\documentclass[aps,twocolumn,superscriptaddress,prd]{revtex4-2}

\usepackage{bm,graphicx,textcomp,amssymb,amsmath,dcolumn,color}
  
\usepackage{multirow}
\usepackage{tabularx}
\usepackage{ctable} 
\usepackage{siunitx}   
\usepackage[utf8]{inputenc}
\usepackage[T1]{fontenc}

\usepackage{bbm}
\usepackage{bbold}   
\usepackage{mathtools}
\usepackage{hyperref}

\usepackage{svg}
\usepackage[margin=8mm,font=small,labelfont=bf, justification=justified]{subcaption}
\usepackage[margin=1mm,font=small,labelfont=bf, justification=justified]{caption}
\renewcommand{\sectionautorefname}{Section}
 
\usepackage[normalem]{ulem}
\usepackage{comment}

\newcommand{\aw}[1]{{\color{black} #1}}

\usepackage{lipsum}  
\usepackage{stackengine}
\newcommand\underlay[4]{%
  \stackengine{0pt}%
  {\kern#2\includegraphics[height=#1]{#4}}%
  {\includegraphics[height=#1]{#3}}%
  {O}{l}{F}{F}{L}%
}
\newcommand\addunderlay[4]{%
  \stackengine{0pt}%
  {\kern#2\includegraphics[height=#1]{#4}}%
  {#3}%
  {O}{l}{F}{F}{L}%
}

\begin{document}

\title{Continuous Dimer Angles on the Silicon Surface:\\ Critical Properties and the Kibble-Zurek Mechanism}

\author{Andreas Weitzel}
\email[]{a.weitzel@hzdr.de}
\affiliation{Helmholtz-Zentrum Dresden-Rossendorf, 
Bautzner Landstra{\ss}e 400, 01328 Dresden, Germany}

\author{Gernot Schaller}
\affiliation{Helmholtz-Zentrum Dresden-Rossendorf, 
Bautzner Landstra{\ss}e 400, 01328 Dresden, Germany}

\author{Friedemann Queisser}
\affiliation{Helmholtz-Zentrum Dresden-Rossendorf, 
Bautzner Landstra{\ss}e 400, 01328 Dresden, Germany}

\author{Ralf Sch\"utzhold}
\affiliation{Helmholtz-Zentrum Dresden-Rossendorf, 
Bautzner Landstra{\ss}e 400, 01328 Dresden, Germany}

\affiliation{Institut f\"ur Theoretische Physik, 
Technische Universit\"at Dresden, 01062 Dresden, Germany}

\date{\today}

\begin{abstract}
	Langevin dynamics simulations are used to analyze the static and dynamic properties of an XY model adapted to dimers forming on Si(001) surfaces. 
    The numerics utilise high-performance parallel computation methods on GPUs. The static exponent $\nu$ of the symmetry-broken XY model is determined to $\nu = 1.04$. The dynamic critical exponent $z$ is determined to $z=2.13$ and, together with $\nu$, shows the behavior of the Ising universality class. For time-dependent temperatures, we observe frozen domains and compare their size distribution with predictions from Kibble-Zurek theory. We determine a significantly larger quench exponent that shows little dependence on the damping or the symmetry-breaking field. \\ 
\end{abstract}

\maketitle
 
\section{Introduction}
	In the area of semiconductor technology, silicon has become the cornerstone material driving the innovations that power our modern world in the \textit{Silicon Age}~\cite{dabrowski2000silicon}.	
	Especially the Si(001) surface of monocrystalline silicon is relevant as it forms an interface with the oxide layer in transistors that isolates silicon nanowires from their environment. While static properties of Si(001), e.g. surface configurations, their energies and electronic structures are thoroughly investigated by theoretical~\cite{ramstad1995theoretical,pillay2004revisit, inoue1994order, ihm1983structural, fu2001molecular} and experimental~\cite{matsumoto2003low, kubota1994streak, brand2023critical, wolkow1992direct, tochihara1994low} works, the dynamic properties~\cite{schaller2023sequential} are not yet well understood. The surface undergoes a continuous order-disorder phase transition with Ising critical exponents~\cite{brand2023critical} between two surface patterns~\cite{tabata1987order}, leading to striking dynamic behavior at the critical point. A phenomenon exhibiting rich dynamics is the Kibble-Zurek mechanism~(KZM)~\cite{kibble1976topology, zurek1985cosmological, zurek1996cosmological, ruutu1996vortex, ulm2013observation, pyka2013topological, laguna1997density, schaller2023sequential, antunes2006domain, bauerle1996laboratory, eltsov2005vortex, volovik2003universe, bunkov2014evolution} which describes the unavoidable non-adiabatic (deviating from instantaneous equilibrium) evolution of systems as they cross phase boundaries. When crossing from a disordered to an ordered phase, the velocity at which systems undergo this transition is directly related to the size of ordered domains, that in turn influences the semiconducting properties~\cite{himpsel1979photoemission, uhrberg1981experimental, handa1989plasma} of the surface. 
 
    The Si(001) surface has often been mapped 
    onto the discrete two-dimensional Ising model~\cite{pillay2004revisit, inoue1994order, ihm1983structural, brand2023critical}.  In this paper, a continuous model of the surface buckling is developed and numerically investigated by Langevin dynamics~\cite{langevin1908theorie, gillespie1996mathematics, frenkel2023understanding} simulations. The emphasis will be put on differences of the dynamics between discrete simulations using the Ising model -- for which a scaling matching the predictions by KZM has been observed~\cite{schaller2023sequential, du2023kibble} --  and continuous modeling using an adapted classical XY model~\cite{kosterlitz2016kosterlitz, kosterlitz1974critical, jose1977renormalization, amit1980renormalisation, barouch1970statistical}. 
 
    This paper is organized as follows: In \autoref{Section::Theoretical-background}, the theory of critical scaling for the static and dynamic case, as well as the Kibble-Zurek scenario, is laid out.  The used model and its mapping to the Si(001) surface is described in \autoref{Section::Modelling-Si}.  The results for the static and dynamic critical properties are discussed in \autoref{Section::Results}.  We conclude our findings in \autoref{Section::Conclusion} and provide technical details in the appendices.
    \begin{figure}
        \centering
        \includegraphics[width=0.8\linewidth]{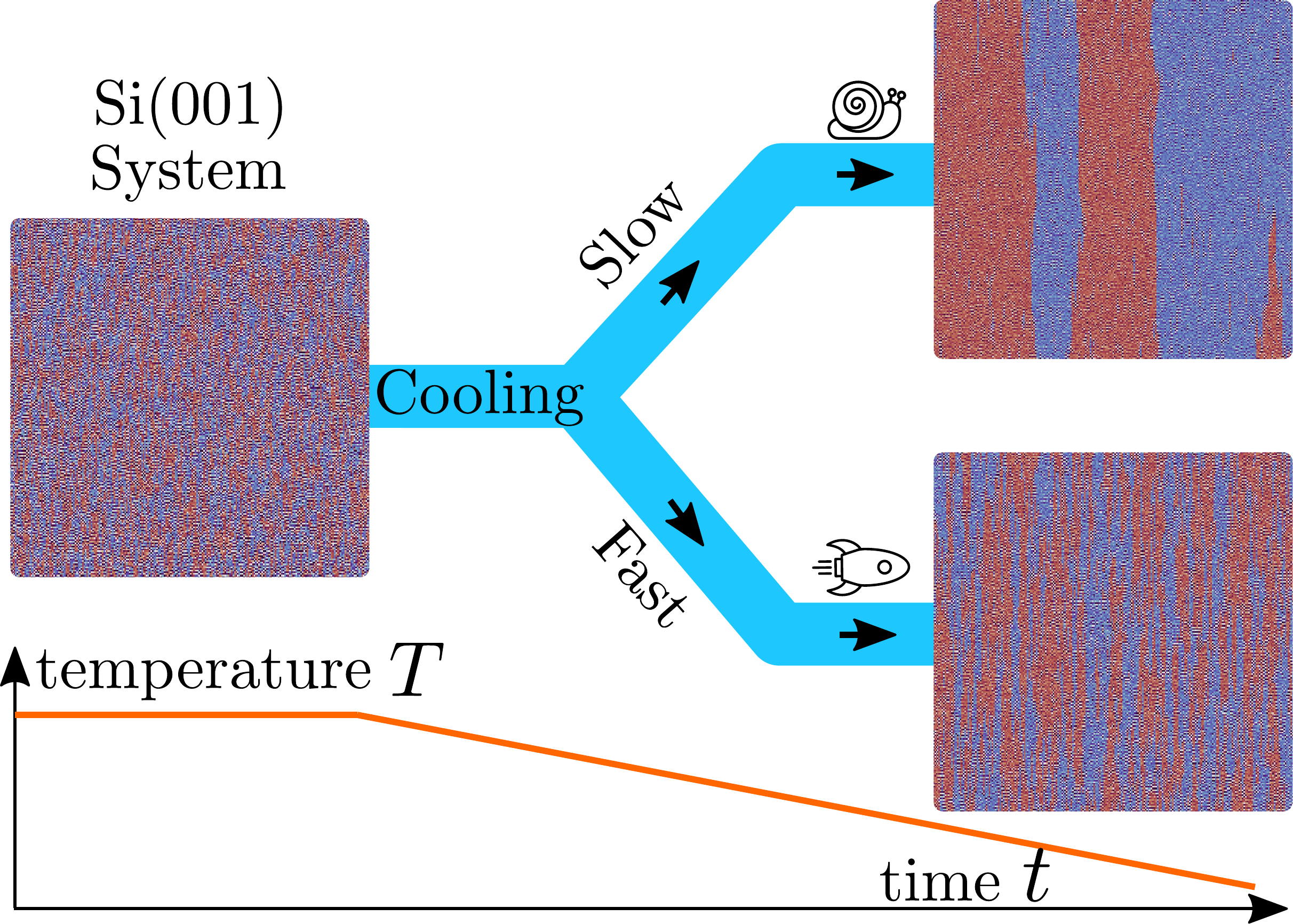}
        \caption{A representation of the silicon (001) surface with the colors mapping into the continuous dimer buckling angle is shown. The system is cooled down with a varying rate. The cooling rate determines the size of ordered patches after the quench.}
        \label{Fig::overview}
    \end{figure}
\section{Theoretical Background} \label{Section::Theoretical-background}
\subsection{Static Scaling}
    Observables of systems near a phase boundary obey universal static scaling laws in the thermodynamic limit. They describe the power law dependence of system quantities in equilibrium on control parameters like the temperature $T$ close to the critical point. For example, the correlation length $\xi$ depends on the reduced temperature 
    \begin{equation}\label{EQ:epsilon}
        \varepsilon =	\frac{T - T_c}{T_c}
    \end{equation} via 
	\begin{equation} \label{Eq::Xi-divergence-amplitude}
		\xi (\varepsilon) = \xi^{\pm} \varepsilon^{-\nu}~,
	\end{equation}
	showing a divergence at the critical value $T =	T_c$ with the \textit{critical exponent} $\nu>0$. This behavior is shared by all systems undergoing a second-order phase transition~\cite{goldenfeld2018lectures, pelissetto2002critical} and is independent of the concrete microscopic properties. The static critical exponent $\nu$ assumes a universal value for systems that share fundamental properties like symmetries or their dimensionality. 
    Systems that possess the same set of critical exponents are said to belong to the same \textit{universality class}. In contrast, the \textit{critical amplitudes} $\xi^\pm$ are non-universal and vary according to the microscopics. The superscript $\pm$ denotes whether the the phase transition is approached from below $(-)$ or above $(+)$ the critical temperature. 
\subsection{Dynamic Scaling}
    In contrast to static scaling, \textit{dynamic scaling} describes the behavior of the relaxation time $\tau$ across the transition. The relaxation time quantifies the time required for fluctuations in the system on the largest length scale to equilibrate.  Near the critical point, it depends on the correlation length as~\cite{hohenberg1977theory}
	\begin{equation} \label{Eq::tau-scaling}
		\tau =	\tau_\xi {\xi^z(\varepsilon)} = \tau_\varepsilon^\pm  \varepsilon^{-\nu z}	~,
	\end{equation}
	defining the universal \textit{dynamic critical exponent} $z$, the critical amplitude $\tau_\xi$ as well as $\tau_\varepsilon^\pm :=	\tau_\xi (\xi^\pm)^z$. Just as the correlation length, the relaxation time becomes infinite at the critical point. This phenomenon is commonly referred to as \textit{critical slowing down}~\cite{hohenberg1977theory, halperin1969scaling, goldenfeld2018lectures} and describes the inability of thermodynamic systems to equilibrate at $\varepsilon\,=\,0$. 
 
    The Kibble-Zurek mechanism~\cite{kibble1976topology, zurek1985cosmological, zurek1996cosmological, ruutu1996vortex, ulm2013observation, pyka2013topological, schaller2023sequential} deals with the question how this non-adiabatic evolution influences the system observables after a quench through the critical point. A simple quench is a decrease of the {reduced} temperature of Eq.~\eqref{EQ:epsilon} linear in time $t$ following
	\begin{equation} \label{Eq::Linear-Quench}
		\varepsilon(t) =	{-} \frac{t}{\tau_Q}~,		
	\end{equation}
    with the \textit{quench timescale} $\tau_Q$. The main statement of the KZM argument is that system quantities after the quench, like $\xi$, are directly proportional to their values $\widehat{\xi}$ at the \textit{freezeout point} defined by
    \begin{equation} \label{Eq::freezeout-point}
        \tau(\,\widehat{t}\,) =	\widehat{t}~,
	\end{equation}
    after which the system equilibration can no longer follow the driving. When approaching the transition point from the high temperature phase, combining Eqs.~\eqref{Eq::tau-scaling},~\eqref{Eq::Linear-Quench} and~\eqref{Eq::freezeout-point} yields the scaling of the frozen correlation length
	\begin{equation} \label{Eq::KZM-scaling}
		\widehat{\xi} = \xi(\varepsilon(\,\widehat{t} \,)) =	\xi^+ / |\varepsilon(\,\widehat{t} \,)|^{\nu} =	\xi^+ \, \bigg|\, \frac{\,\tau_Q\,}{\tau_\varepsilon^+}\, \bigg|^{\mu_{\text{KZM}}}~
	\end{equation}
    with
    \begin{equation} \label{Eq::KZM-Exponent}
        \mu_{\text{KZM}} = {\frac{\nu}{1 + \nu z}} 
    \end{equation}
	being the \textit{quench exponent} according to KZM. Equation~\eqref{Eq::KZM-scaling} quantifies the dependence of the quenched correlation length on the quench timescale $\tau_Q$. A specialty of the KZM argument is that it enables to predict the behavior of quenched systems solely from universal exponents. 
\section{Model}
\label{Section::Modelling-Si}
	\begin{figure*}
    	\centering
    	\begin{subfigure}{0.26\textwidth}
    		\centering
    		\includegraphics[width=0.9\linewidth]{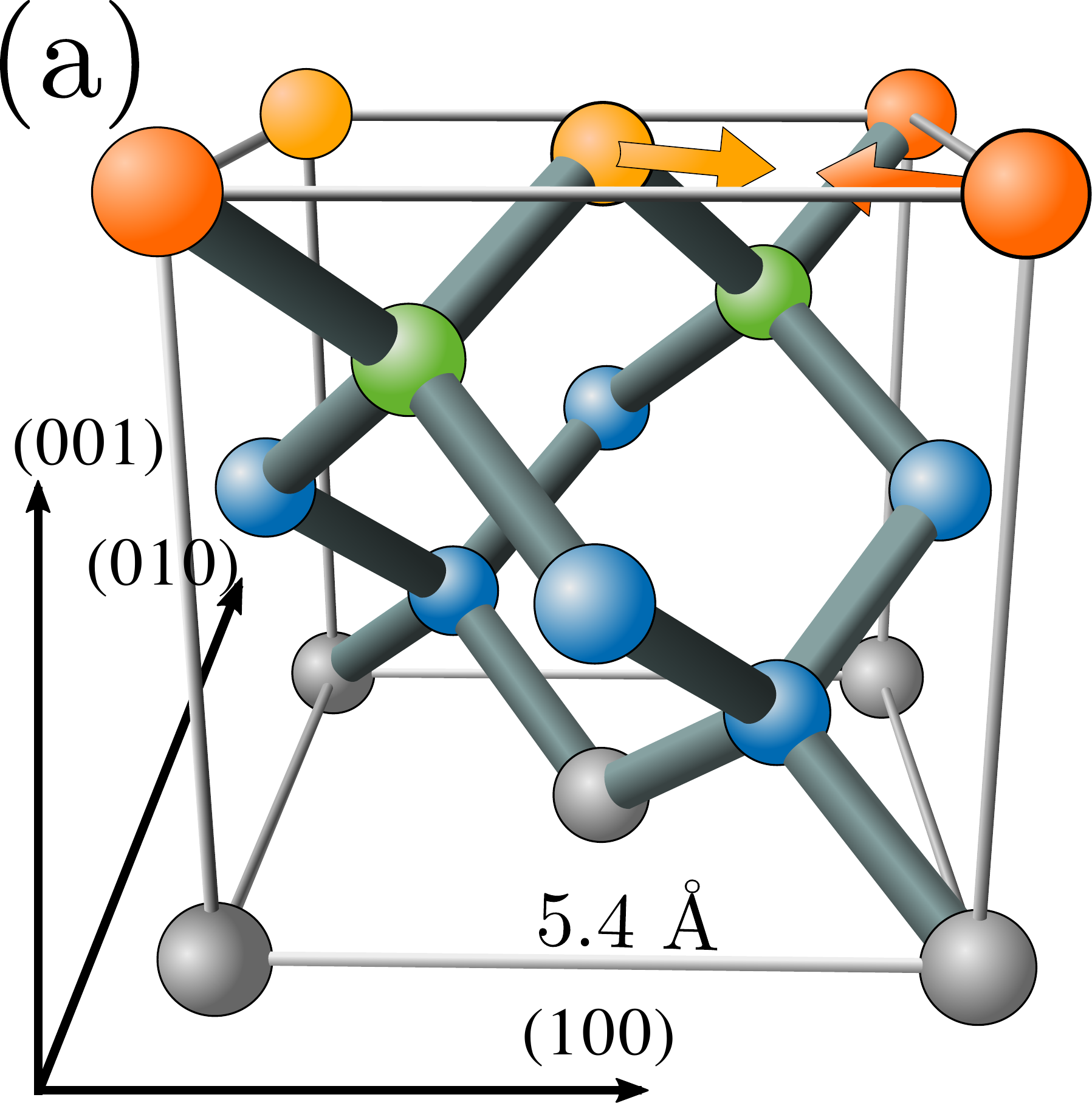}
    		\label{Fig::SiliconDiamond}
    	\end{subfigure}
    	\begin{subfigure}{0.35\textwidth}
    		\centering
    		\includegraphics[width=0.9\textwidth]{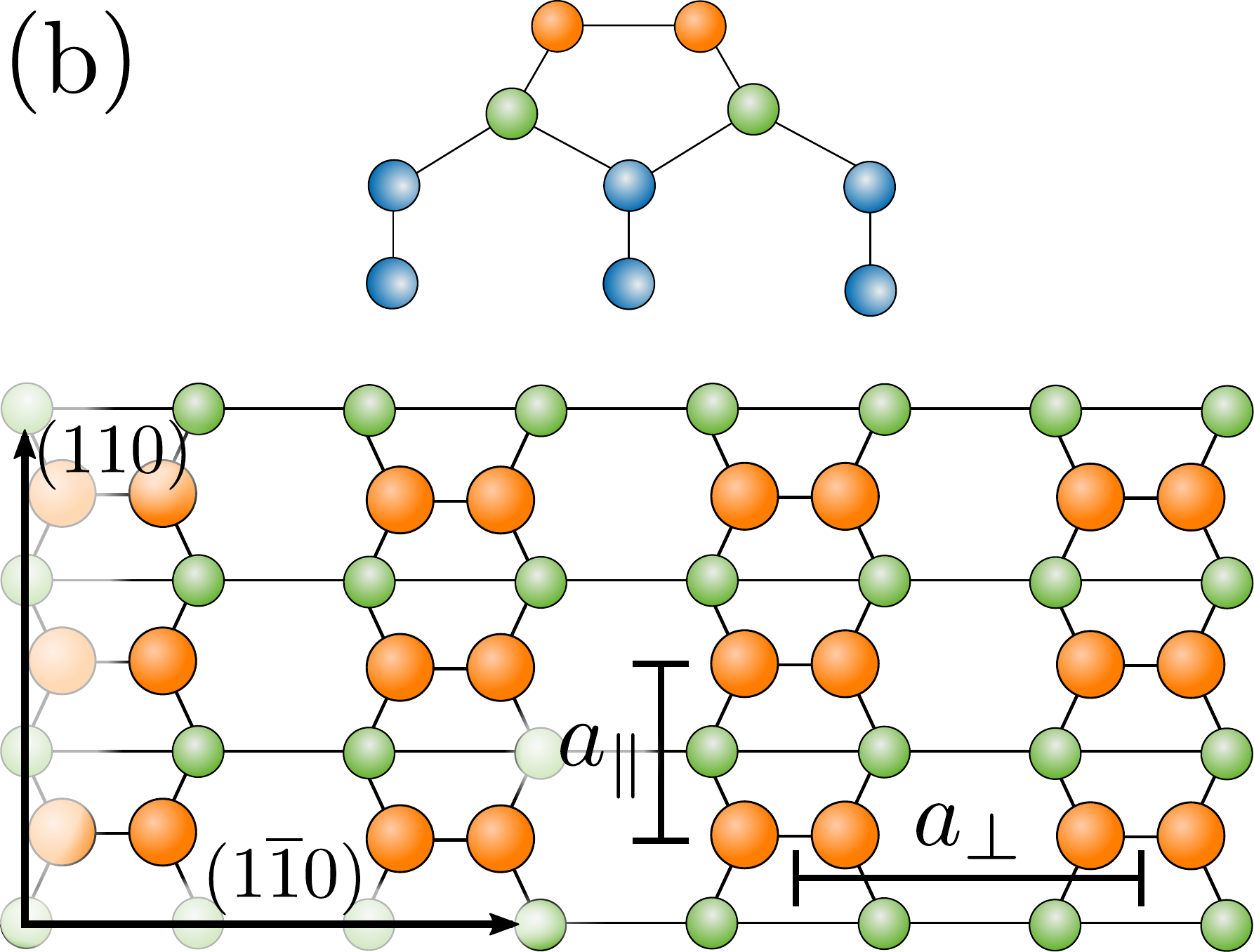}
    		\label{p(2x1)-symmetric}
    	\end{subfigure}
    	\begin{subfigure}{0.35\textwidth}
    		\centering
    		\includegraphics[width=0.9\textwidth]{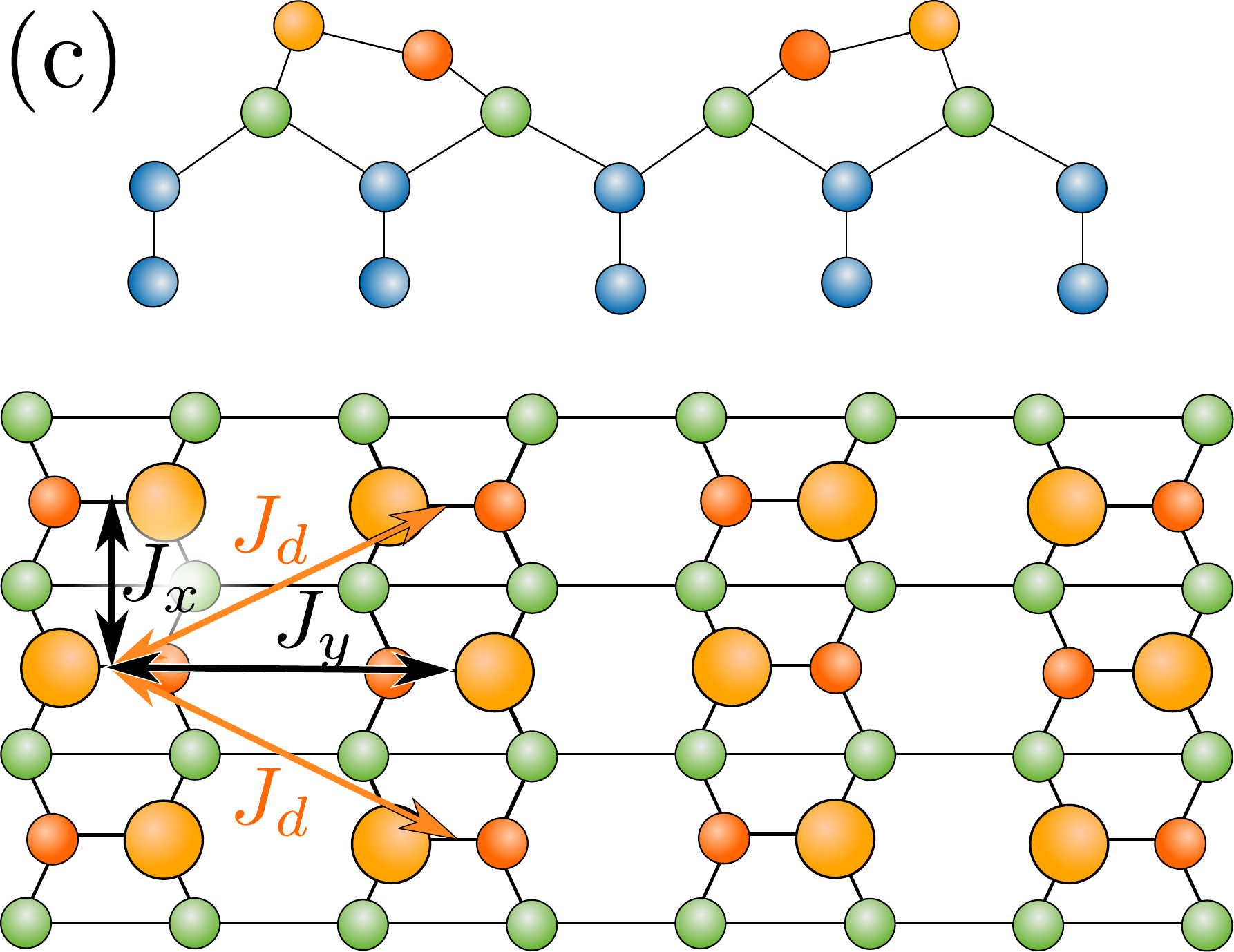}
    		\label{c(4x2)}
    	\end{subfigure}		
    	\caption{The figure, inspired by~\cite{brand2023critical}, shows the crystal structure of silicon and two surface configurations. The coloring of the silicon atoms corresponds between (a) and {(b), (c)}. In the following, the surfaces are addressed using Wood's notation for overlayers~\cite{attard1998surfaces}. \textbf{(a)} The crystalline structure of Si in its solid state is shown~\cite{sistrucure}.  The orange silicon atoms dimerize during reconstruction of the (001) surface. The coordinate axes are denoted by their Miller indices and normal to their corresponding crystallographic planes. In (b) and (c), Si(001) surface patterns are shown from a side and top-down perspective. In the top-down view, silicon atoms that are closer to the reader are magnified. \textbf{(b)} The formation of dimers results in the symmetric $p(2\times1)$ reconstruction and reduces the energies by $1.8~\text{eV}$ per dimer compared to the non-dimerized structure. \textbf{(c)} The dimers are unstable to vertical buckling. The buckling pattern that was found to have the lowest surface energy is the $c(4\times2)$ reconstruction. The fundamental couplings $J_\delta, \delta \in \{x, y, d\}$ are indicated by arrows.}
    	\label{Fig::dimer-configs}
	\end{figure*}	
    The upcoming investigations will make use of molecular dynamics (MD)~\cite{frenkel2023understanding}, or more precisely, \textit{Langevin dynamics} simulations~\cite{langevin1908theorie, gillespie1996mathematics}. MD simulations numerically solve Newton's equations of motion to describe the behavior of many-particle systems, thus going beyond a discrete modeling of the dimers, which omits the details of the dynamics. Therefore, it is desirable to investigate whether a simulation with a continuum of states results in other dynamic critical behavior than the discrete models. Below, we first introduce a model similar to the Ising model that offers continuous states and reproduces the experimentally observed Ising static critical exponents and explain its relation to the silicon surface as well as its numerical Langevin dynamics simulation.
\subsection{Classical XY Model} \label{Section::XY-Model}
    The classical two-dimensional XY model (or planar rotor model) without symmetry-breaking fields is characterized by sites each hosting a unit-length rotor in two dimensions $\vec{s} =	( \cos (\vartheta), \sin (\vartheta))$, with $\vartheta \in [0, 2 \pi )$ characterizing the angle of the in-plane rotors. Its nearest-neighbor Hamiltonian on a lattice with sites $(i,j)$ is given by 
    \begin{align} \label{Eq::generic-XY}
        \tilde{H}_\text{XY} = - J_\parallel \sum_{ij} \vec{s}_{i,j}\cdot \vec{s}_{i+1,j} - J_\perp \sum_{ij} \vec{s}_{i,j} \cdot \vec{s}_{i,j+1}\,,
    \end{align}
	with the coupling constants $J_\delta$ and $\delta \in \{ \perp, \, \parallel \, \}$. In contrast to the two-dimensional Ising model~\cite{onsager1944crystal, mccoy1973two}, an exact closed solution to the two-dimensional XY model is unknown. However, it has been studied in numerous analytical~\cite{kosterlitz2016kosterlitz, kosterlitz1974critical, jose1977renormalization, amit1980renormalisation, barouch1970statistical} and numerical~\cite{hasenbusch2005two, janke1993high, edwards1991multi}  works. Without symmetry-breaking fields, it does not exhibit an ordinary phase transition with universal critical exponents, but undergoes the Kosterlitz-Thouless transition~\cite{kosterlitz1974critical}. In this case, the system transitions from an unordered state to a quasi-ordered state of vortex-antivortex pairs. The correlation length still diverges, but does not follow Eq.~\eqref{Eq::Xi-divergence-amplitude} anymore. Instead, \aw{when approaching the transition from the disordered state, }$\xi$ diverges exponentially and\aw{, in the ordered state, the spin-spin correlations} vary continuously with the temperature \cite{kardar2007statistical}. 
 
    The quantum version of Eq.~\eqref{Eq::generic-XY} can be obtained by replacing the spins with Pauli matrices~\cite{lieb1961two}. It is relevant for insulator to superfluid quantum transitions~\cite{pazmandi1995quantum}, effective interactions between quantum dots~\cite{imamog1999quantum}, the quantum-Hall effect~\cite{wang2001entanglement}, as well as cavity QED systems in quantum computers~\cite{wang2001entanglement}. In two dimensions, it also undergoes a Kosterlitz-Thouless transition~\cite{harada1998kosterlitz, ding1990kosterlitz}, like its classical counterpart.
    
    \aw{We generalize Hamiltonian~\eqref{Eq::generic-XY} by} a $p$-fold symmetry-breaking field of strength $h$ \aw{as well as a multiplying factor $q$ (similar to the generalized XY model \cite{poderoso2011new})} to  to obtain
    \begin{align} \label{Eq::XY-Hamilton-Field}
			H_{\text{XY}} =&- J_\parallel \sum_{i,j}^{}   \cos \big( \aw{q} \left(\vartheta_{i,j} - \vartheta_{i+1, j}\right) \big)\nonumber\\
   &- J_\perp \sum_{i,j} \cos \big( \aw{q} \left(\vartheta_{i,j} - \vartheta_{i, j+1} \right) \big) \nonumber \\
			&+ h \sum_{i,j} \cos(p\vartheta_{i,j})\,.
	\end{align} 
    \aw{The purpose of $q$ is to enable state space restrictions.} The critical properties change drastically if a symmetry-breaking field is introduced. In this case, José et al.~\cite{jose1977renormalization} showed by renormalization group calculations that the transition changes to the universality class of the $p$-state \aw{clock model or vector Potts model}~\cite{potts1952some}. The determining factor for the specified universality is the number of minima of the symmetry-breaking term. The case $p=2$ yields the Ising transition for any non-zero $h$.
\subsection{The Si(001) Surface}
    \begin{figure*}
    	\centering
    	\begin{subfigure}[t]{0.70\textwidth}
        	\centering
	        \includegraphics[width=0.9\linewidth]{State-Mapping.pdf}
        \end{subfigure}
       	\begin{subfigure}[t]{0.28\textwidth}
    		\centering
    		\includegraphics[width=0.9\linewidth]{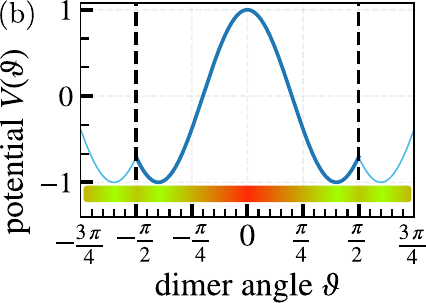}
	      	\label{Fig::Potential}
    	\end{subfigure}
        \caption{\aw{\textbf{(a)}} Unit circles are shown as visualizations of the state space. The colored part of the unit circle represents the allowed states. The states are depicted as arrows inside the unit circle. The buckling angles of the silicon dimers are mapped to the rotors of the XY model. Since the silicon atoms are indistinguishable, a rotation by $\pi$ maps to the same state. Hence, the state space is only half the unit circle. The coloring of the allowed states corresponds to the value of the potential $V(\vartheta)$. The vertical dimer position ins highly unlikely and the minimum of the potential is roughly located at the equilibrium buckling angles of the dimers. Note that the left and right vectors are the same state and one of them \aw{is} excluded by defining $\vartheta \in [-\pi / 2, \pi / 2)$. \textbf{\aw{(b)}} \aw{The on-site potential is shown in dependence of the dimer angle. The allowed phase space is delimited by vertical dashed lines. Outside the allowed states the angles are mapped back to $\vartheta \in [-\pi / 2, \pi / 2)$ by modulo operation. The periodic continuation of the potential is indicated by the light blue line.} }
        \label{Fig::States}
	\end{figure*}
    Silicon crystallizes in a diamond cubic structure, as shown in \autoref{Fig::dimer-configs} (a). When cutting this crystal structure along the crystallographic (001) plane, the resulting surface reconstructs by dimerization of the surface atoms~\cite{chadi1979atomic} (see \autoref{Fig::dimer-configs} (b)). The formation of dimers lowers the surface energy by roughly $1.8~\text{eV}$ per dimer~\cite{ramstad1995theoretical, batra1990atomic, dabrowski1992self}.	The dimers can further reduce their energy by buckling vertically. They tilt to an angle of about $18^\circ$ with the surface plane~\cite{ramstad1995theoretical, pillay2004revisit}, which lowers the surface energy by another $0.15~\text{eV}$~\cite{inoue1994order} per dimer. A charge transfer of approximately $0.1 e$~\cite{brand2023critical, landemark1992core} is induced by the buckling.
    
	Theoretical~\cite{ramstad1995theoretical, pillay2004revisit, inoue1994order, brand2023critical} and experimental (low energy electron diffraction~\cite{matsumoto2003low, kubota1994streak, brand2023critical} and scanning tunneling microscopy~\cite{wolkow1992direct, tochihara1994low}) investigations have found the $c(4\times2)$ reconstruction, shown in \autoref{Fig::dimer-configs} (c), to be the lowest energy geometry. It minimizes the interaction energy as well as the surface stress~\cite{pillay2004revisit}. The alternating buckling in both directions suggests an antiferromagnetic interaction along both directions. However, apart from an antiferromagnetic interaction along the dimer rows, experimental observations~\cite{brand2023critical,brand2023dimer} actually indicate ferromagnetic transverse couplings $J_y~>~0$ and additional ferromagnetic diagonal couplings $J_d~>~0$. The dimer interactions are strongly anisotropic, with $|J_x|$ being much larger than $|J_y|$, enforcing alternating buckling in $(110)$ direction. The ferromagnetic diagonal interactions $J_d$ overpower $J_y$, so that diagonal alignment is preferred, which in turn implies anti-alignment in $(1\overline{1}0)$ direction, so that the effective transverse coupling is also antiferromagnetic. Because of the strong anisotropy, it is satisfactory to consider only effective nearest neighbor couplings, which enables us to absorb the diagonal interactions $J_d$ into a an effective transverse 
    \begin{equation}
        J_\perp = J_y - 2 J_d <0
    \end{equation}
    across the dimer rows~\cite{brand2023dimer}. The effective coupling along the dimer rows remains antiferromagnetic $J_\parallel = J_x$.
 
    The silicon (001) surface exhibits an order-disorder phase transition from the disordered $p(2\times1)$ phase (\autoref{Fig::dimer-configs} (b)) to the ordered $c(4\times2)$ (\autoref{Fig::dimer-configs} (c)) reconstruction at a critical temperature of about $T_c \approx 200~\text{K}$~\cite{tabata1987order, brand2023dimer}. The $p(2\times1)$ structure is short term for the disordered phase since fast flipping of the dimers at a rate of about $ 10^{11} \text{ Hz}$~\cite{dabrowski1992self} let the system appear to be in the $p(2\times1)$ state at high-temperature measurements. This continuous phase transition will be of central importance in the following discussion.
 
	Below $T_c$, the strong anisotropy leads to long streaks of order along the dimer rows and short domains of order across the dimer rows. The correlation length amplitude in parallel direction $\xi_\parallel^+ $ is larger than the transverse amplitude $\xi_\perp^+$ by a factor of five~\cite{brand2023dimer}. 	In units of the lattice constants $a_\parallel$ and $a_\perp$ (see \autoref{Fig::dimer-configs}~(b)) this implies $(\xi_\parallel^+ / a_\parallel) \big / (\xi_\perp^+ / a_\perp) \approx 10$. Si(001) was mapped onto the exactly solvable two-dimensional  Ising model in previous research~\cite{brand2023dimer, pillay2004revisit, ihm1983structural, schaller2023sequential, inoue1994order} by assigning its two discrete states to the two equilibrium buckling angles of the dimers. A good agreement between the experimentally measured static critical properties of the surface's phase transition and the Ising universality class has been found~\cite{brand2023critical, brand2023dimer}. Furthermore, simulations based on kinetic Monte Carlo methods~\cite{bortz1975new} have revealed a quench scaling that roughly matches the KZM~\cite{schaller2023sequential}.
\subsection{Mapping the XY Model to Si(001)}	
    The natural choice of mapping the Si(001) dimers to the XY model is to identify the dimer buckling angles with the rotor angles $\vartheta$. Since silicon atoms are indistinguishable, a rotation of the dimer by $\pi$ results in the same state. We will therefore restrict the state space of the rotor to $\vartheta \in [-\pi/2, \pi/2)$. \aw{To ensure that the interaction terms in Eq.~\eqref{Eq::XY-Hamilton-Field} have the same periodicity as $\vartheta$, we set $q=2$ inside the cosine terms.} The correspondence between the rotors and the dimers is depicted in \autoref{Fig::States}. 
    The parameter $p$ in \eqref{Eq::XY-Hamilton-Field} determines the number of minima of the on-site potential and thereby the equilibrium position of the dimers. As we have cut the state space in half, the effective integer $p_{\text{eff}}$ relevant for the \aw{clock model} classification of \autoref{Section::XY-Model} will be $p_{\text{eff}} = \lceil p / 2 \rceil$. Since the dimers have two equilibrium buckling angles $\vartheta_\pm$, we demand $2 < p \leq 4$, corresponding to an effective $p_{\text{eff}} = 2$. Due to the additional antiferromagnetic interaction, the equilibrium position of the dimers will be somewhere between the minima of the symmetry-breaking potential and $ \pm \pi /	4  \leq \vartheta_\pm $. Since the upcoming investigations will not involve quantitative predictions but rather an assessment of universal behavior, we will set $p =	2.5$ and leave the exact calibration of equilibrium buckling angles with experimental data to future work.
\subsection{Langevin Dynamics}
    A molecular dynamics approach solving the
    equations of motion for every dimer will be employed to numerically investigate the adapted XY model. The Si(001) surface is thermally coupled to the silicon bulk, which acts as a thermal reservoir. One way to model thermalization with this reservoir, even for interacting particle potentials, is to replace Newtons equations of motion with Langevin equations. In other words, the movements of the dimers will be modeled as Brownian motion in a (double-well) potential. Since we want to describe the full dynamics of the silicon dimers in terms of the buckling angle, we add the kinetic energy to the Hamiltonian of Eq.~\eqref{Eq::XY-Hamilton-Field}, reading
	\begin{equation} \label{Eq::XY-Hamilton-Kinetic}
		H =	\sum_{i,j} \frac{1}{2} I \omega_{i,j}^2 + H_\text{XY},
	\end{equation}
	with the dimer moment of inertia $I$ and the angular velocity $\omega_{i,j}$. 
 
    In the following, we will work with a dimensionless representation of the problem, meaning that all quantities, including time $t$ or angular velocity $\omega$, will be given in multiples of $I$
    or $I^{-1}$, respectively.
 
    The dimensionless stochastic equations of motion for rotary motion in a (periodic) potential $V(\{\vartheta\})$ are, in units of $I$, given by~\cite{gillespie1996mathematics, risken1996fokker}
	\begin{align}
		&\frac{\text{d}}{\text{d}t} \vartheta_{i,j}(t) =	 \omega_{i,j}(t)~, \label{Eq::Si-Langevin-theta} \\
		&\frac{\text{d}}{\text{d}t} \omega_{i,j}(t) =	- {\eta} \omega_{i,j}(t) - \frac{\partial V(\{\vartheta\})}{\partial \vartheta_{i,j}} + \sqrt{2 \eta T }\, \Gamma(t)~, \nonumber \label{Eq::Si-Langevin-omega}
	\end{align}
	with the \textit{damping} $\eta$, temperature $T$, and the \textit{Gaussian white noise process} $\Gamma(t)$~\cite{gillespie1996mathematics}. The latter satisfies 
		\begin{equation} \label{Eq::Gauss-dist-property}
		\left \langle \Gamma(t) \right \rangle = 0 \qquad \text{and} \qquad  			\left \langle \Gamma(t) \Gamma(t + t')\right \rangle =	\delta(t')~.
	\end{equation}
	\aw{To evaluate the on-site potential, we map the angles back to the allowed interval $\vartheta_{i,j} \in [-\pi/2, \pi/2)$ by modulo operation after every integration step.} The simulation is implemented in CUDA C\texttt{++}~\cite{cuda} using Thrust~\cite{thrust} as high level interface. The integration scheme is described in \renewcommand{\sectionautorefname}{Appendix} \autoref{Section::Integration-Method}. We have benchmarked the thermalization of this system for dimer pairs in \autoref{Section::Benchmarks}. Additionally, the dynamics of Brownian motion in a harmonic potential and the used integration step size $\text{d}t$ have been tested. \renewcommand{\sectionautorefname}{Section}
\section{Results}	\label{Section::Results}
	\begin{figure*}
		\begin{subfigure}[t]{0.65\textwidth}
			\includegraphics[width=1\linewidth]{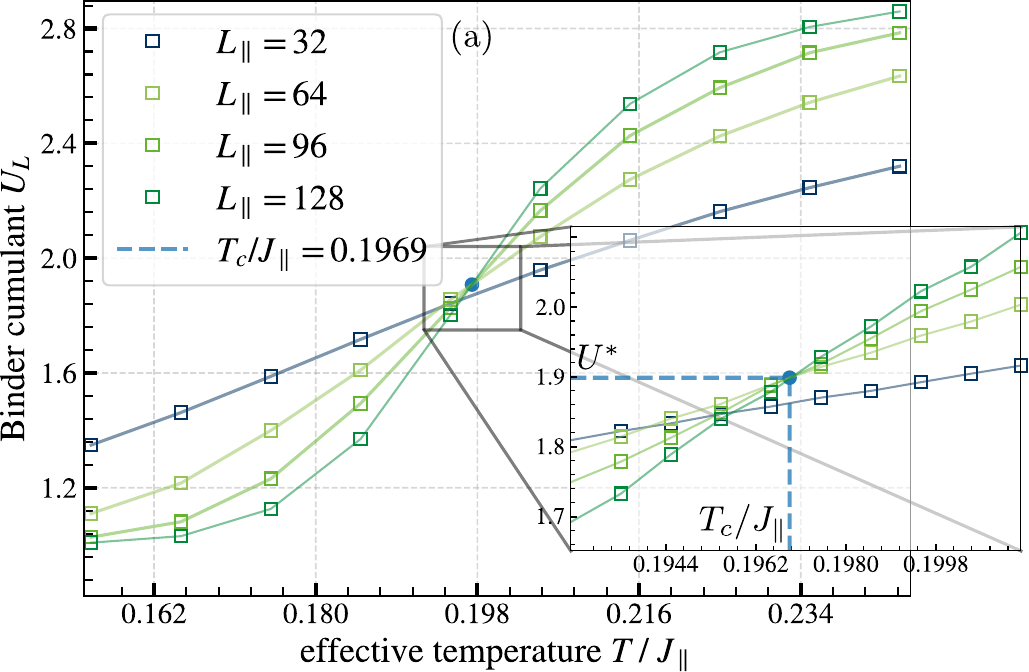}
		\end{subfigure}
		\begin{subfigure}{0.3\textwidth}
			\begin{subfigure}{\textwidth}
				\includegraphics[width=0.95\linewidth]{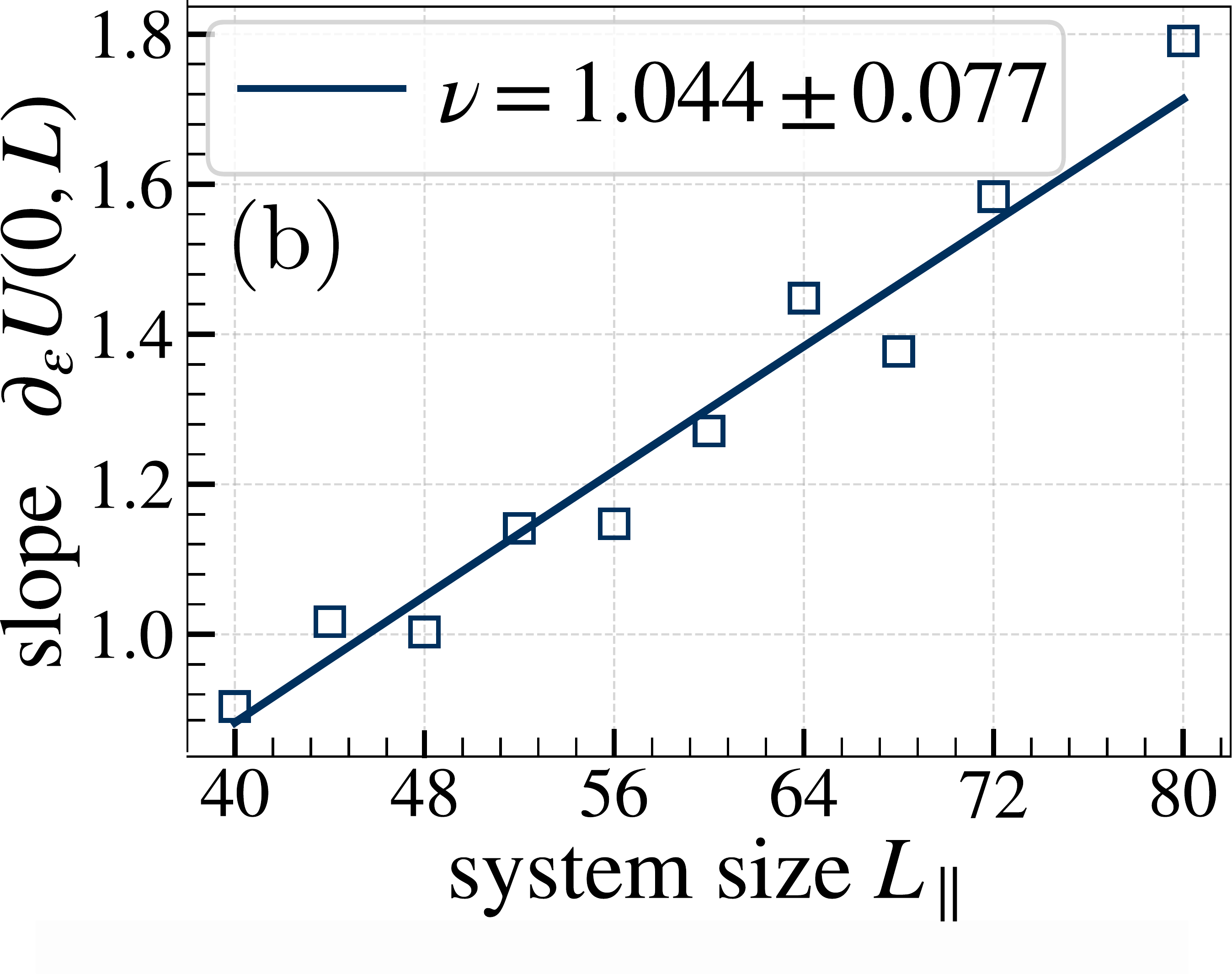}
			\end{subfigure}
			\begin{subfigure}{\textwidth}
				\includegraphics[width=0.95\linewidth]{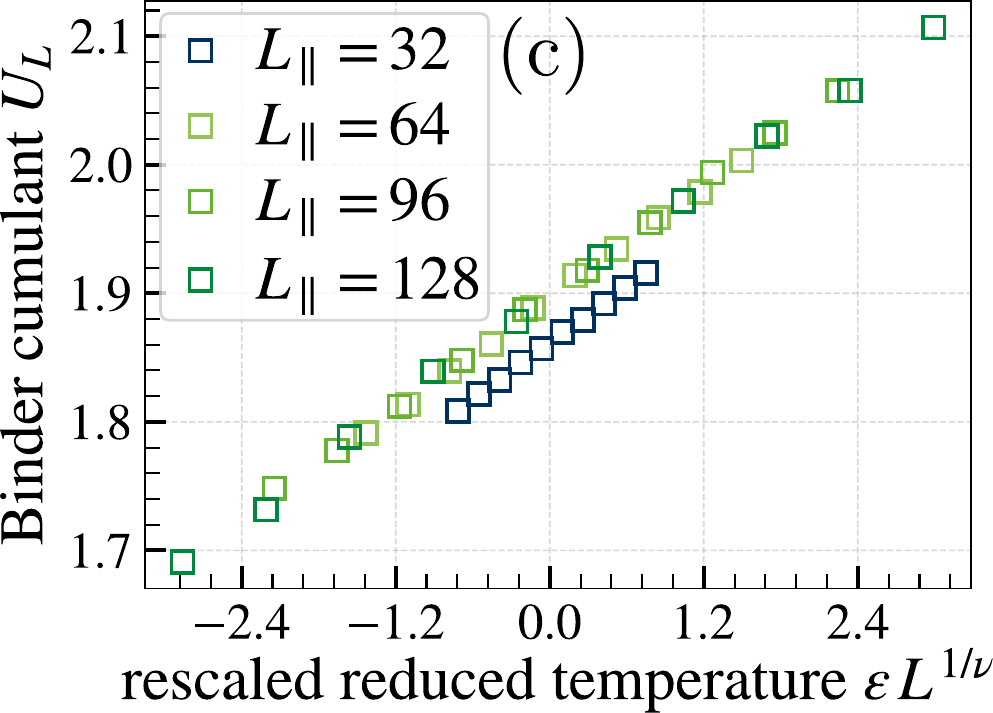}
			\end{subfigure}
		\end{subfigure}
		\caption{\textbf{(a)} The results of ${U}(\varepsilon, L)$ for selected sizes $L$ are computed close to the critical point. The systems were initialized in a totally ordered state with $U(0, L) = 1$ as it minimizes the time of relaxation to $U(\varepsilon, L)$~\cite{binder2022monte}. Except for the smallest $L=32$ all cumulants intersect approximately at the same point $(\beta_c J_\parallel)^{-1} = 0.1969$. \textbf{(b)} The derivatives ${\partial_\varepsilon U(\varepsilon, L)} \propto L^{{1}/{\nu}}$ are shown in dependence of the system size. They are calculated using a standard central difference method including, among others, the data points of (a). The fitting yields $\nu = 1.04 \pm 0.08$ (fit uncertainty~\cite{vugrin2007confidence}) which is in good accordance with the Ising model. \aw{\textbf{(c)} Close to the critical point, the Binder cumulant for different system sizes collapses on a line if plotted versus $\varepsilon L^{1/\nu}$, like expected from Eq.~\eqref{Eq::Binder-cum-FSS}. Corrections to FSS are again visible for $L=32$.} } 
		\label{Fig::Binder-Cum-Result}
	\end{figure*}
	The simulations have been performed with a coupling constant ratio of $J_\parallel / J_\perp = 100$ as the numerics have shown that this choice reproduces the experimental equilibrium correlation length ratio~\cite{brand2023dimer} of  $(\xi_\parallel^+ / a_\parallel) \big / (\xi_\perp^+ / a_\perp)~\approx~10$. To keep all values around unity, $I J_\parallel = 10$ and $I J_\perp = 0.1$ are chosen. Unless marked otherwise, the damping is set to $\eta =	1$. The damping does not influence the static properties of the system and the effects on the dynamics are investigated later. Since the energy barrier between the equilibrium positions is proportional to $h$, the magnitude of $h$ slows down the dynamics of the system significantly. As the computational resources are limited, the selected value of $h$ is not fitted to the experimental system but set to $I h = I \sqrt{J_\perp J_\parallel} = 1$.
\subsection{Static Scaling} \label{Section::Static-scaling}
    Since some modifications have been made to the anisotropic XY model, e.g. restricting the state space and using rational $p$ instead of the integer values used in~\cite{jose1977renormalization}, it is desirable to verify that the model still belongs to the expected Ising universality class. We therefore start by extracting the static critical exponent $\nu$. The solution of the equations of motion is computationally expensive so that finite-size scaling (FSS) methods~\cite{goldenfeld2018lectures, domb1983vol8} will be employed. More specifically, we will make use of the FSS of the Binder cumulant~\cite{binder1981finite} defined by
	\begin{equation} \label{Eq::Def-Binder-Cum}
		U(\varepsilon, L) = \frac{\langle M^4(L) \rangle_\varepsilon}{\langle M^2(L) \rangle^2_\varepsilon}~,
	\end{equation}
	with the ensemble average $\langle \, \cdot \, \rangle_\varepsilon$ at reduced temperature $\varepsilon$ and $M(L)$ being the lattice average 
	\begin{equation} \label{Eq::lattice-average}
		M(L)	=	\frac{1}{{N(L)}} \sum_{i=0}^{N(L)} s_i
	\end{equation}
	of a system of size $L$ with $N(L)$ lattice sites, e.g. $N(L) = L^2$ for a two-dimensional quadratic lattice. In our adapted XY model, the site's state $s_i$ is characterized by $s_i(\vartheta_i) = \sin(\aw{q}\vartheta_i)$ with \aw{$q =	2$} so that our order parameter is analogous to the magnetization in $x$-direction in the model with full state space $M_x = \left\langle \sin(\vartheta_i) \right\rangle$. Additionally, the factor two enforces the $\pi$-periodicity of our model and roughly maps the equilibrium buckling angles $\vartheta_\pm$ to $\sin(2\vartheta_\pm) \approx \pm 1$, like in the Ising model. The finite-size scaling of $U(\varepsilon, L)$ can be parameterized~\cite{pelissetto2002critical, campostrini2001critical} as
	\begin{equation} \label{Eq::Binder-cum-FSS}
		U(\varepsilon, L) =	U^* + A \varepsilon L^{1/\nu} \left(1 + B L^{-\tilde{\omega}} + ...\right)~,
	\end{equation}
	with $\tilde{\omega}$ being the smallest irrelevant exponent in renormalization group language~\cite{goldenfeld2018lectures, pelissetto2002critical}. The symbols $A$ and $B$ are unknown proportionality constants. The curves $U(\varepsilon, L)$ intersect for different system sizes $L$ at $U(0, L) = U^*$. For Ising lattices where $s_i=\sigma_i$, $\sigma_i \in \lbrace -1, 1 \rbrace$, and thus $M$ is the magnetization per spin, it can be shown~\cite{binder1981finite} that the Binder cumulant jumps from its zero-temperature limit  $U(\varepsilon < 0, \infty)=1$ below the critical temperature to its infinite-temperature limit $U(\varepsilon >0, \infty) =	3$ above the critical temperature. Although the Binder cumulant $U$ depends on factors like boundary conditions, lattice shapes and anisotropies~\cite{selke2006critical}, the intersection remains at $T_c$ regardless of these factors and can therefore be used to extract the critical temperature. The preceding implications can be generalized to spin dimensionalities larger than one~\cite{binder1981critical} and have been shown to be consistent with the XY model~\cite{landau1983non, bernreuther1988investigation, loison1999binder}. By simulating systems around the critical point and computing $\partial_\varepsilon U(\varepsilon, L)$ at $\varepsilon = 0$, the exponent $\nu$ may then also be calculated by fitting  
	\begin{equation} \label{Eq::FSS-dU_dT}
		\ln \left(\frac{\partial U(\varepsilon, L)}{\partial \varepsilon}\bigg |_{\varepsilon = 0} \right) \approx	\ln \left(A L^{1/\nu} \right) =	\ln (A) + \frac{1}{\nu} \ln (L)
	\end{equation}
	and extracting the slope $1 / \nu$. \autoref{Fig::Binder-Cum-Result} shows excellent agreement with the Ising critical exponent $\nu =	1$. The evaluation and assessment of the simulation is described in \renewcommand{\sectionautorefname}{Appendix} \autoref{Section::observable-measurement} \renewcommand{\sectionautorefname}{Section}.
\subsection{Dynamic Scaling} \label{Section::dynamic-scaling}
\begin{figure}
		\centering
		\includegraphics[width=0.95\linewidth]{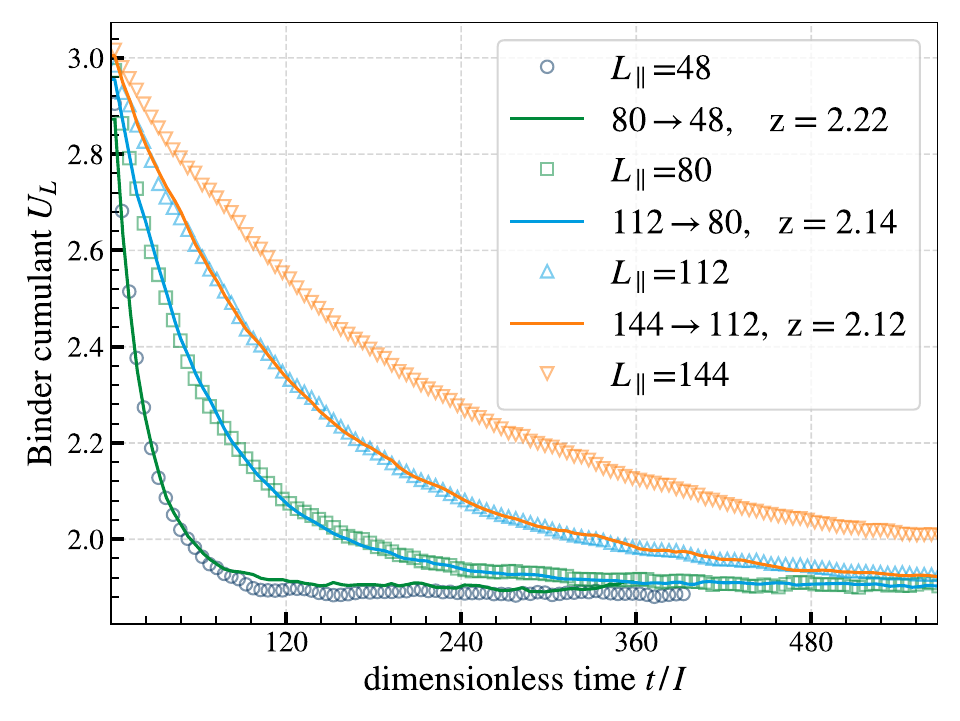}
		\caption{The relaxation of the Binder cumulant is shown versus the dimensionless time for selected sizes $L$. The dynamic critical exponent $z$ can be extracted by finding a time rescaling factor $ b^{-z} = ({L_1}/{L_2})^{-z}$ such that $U(t, \varepsilon, L_1)$ and $U(b^{-z}t, \varepsilon, L_2)$ overlap.	The best rescaling of $U(t, \varepsilon, L_1) \rightarrow U(t, \varepsilon, L_2)$ is selected by rasterizing $z$ in steps of $0.01$ and minimizing the squared error between the interpolated curves. The rescalings $144 \rightarrow 112$ and $112 \rightarrow 80$ yield an average exponent of $\texttt{}z = 2.13$, which is very close to the Ising universality class $z=2.16$ (see \autoref{Section::Ising-Crit-Exp}).}
		\label{Fig::Cum-Relax-z-extrac}
	\end{figure}

	Additionally, the Binder cumulant is useful to extract the dynamic critical exponent $z$. We employ the cumulant relaxation method introduced by Li et al. in~\cite{li1995dynamic} and extend the cumulant by a time dependence $U(\varepsilon, L)~\rightarrow~U(t, \varepsilon, L)$. Their argument exploits that a totally unordered system at $U(0, \infty, L) = 3$ coupled to a reservoir at $\varepsilon =	0$ relaxes to its equilibrium value $U(\infty, 0, L) =	U^*$. The time scale at which this relaxation happens increases with the system size. Quantitatively, it can be shown that the time resolved behavior of systems of different sizes $L_1$ and $L_2$ can be formulated as \cite{li1995dynamic}
	\begin{equation}
		U(t, 0, L_1) =	U(b^{-z} t, 0, L_2)~,
	\end{equation}
	with $b =	L_1 / L_2$ being the spatial rescaling factor. The dynamic exponent can be extracted by finding a temporal rescaling factor $b^{-z}$ so that the relaxation curves for different sizes $L_i$ collapse. Since this only holds precisely for the critical relaxation, the critical temperature of the system has to be determined beforehand by using the intersection in \autoref{Fig::Binder-Cum-Result} (a). The exponent $z$ has then been calculated by minimizing the squared deviation between rescaled curves for a discrete rasterization of $z$-values with a step size of $0.01$. \autoref{Fig::Cum-Relax-z-extrac} shows that the extracted $z$ is very close to the best guess of the Ising $z$ in \autoref{Tab::Ising-z-values}. Hence, we believe that our model still belongs to the Ising universality class.
	\renewcommand{\sectionautorefname}{Appendix}
	\renewcommand{\sectionautorefname}{Section}
    \begin{figure*}
    	\centering
    	\includegraphics[width=0.79\linewidth]{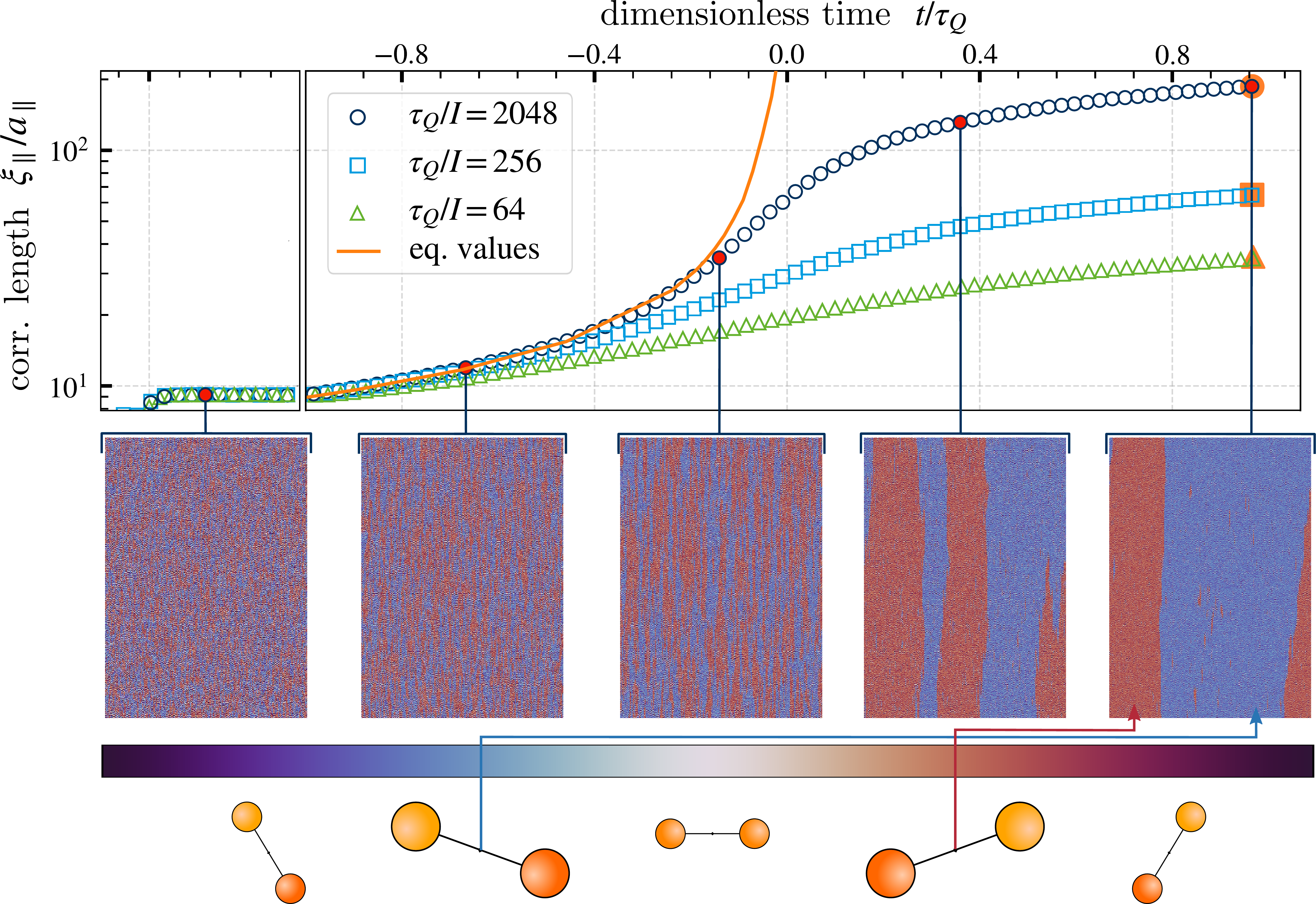}
    	\caption{The time resolved parallel correlation length $\xi_\parallel(t)$ is shown for quenches of different timescales in the top plot. The systems go through an equilibration phase before the quench starts. The orange curve depicts the equilibrium values $\xi_\parallel(\varepsilon(t))$, visualizing where the systems deviate from adiabatic evolution. For the red markers, section snapshots of one representative quenched system at the specified time are shown below. The coloring of the meshs map into the dimer angle and is depicted by the cyclic colorbar on the bottom. In the frozen state after the quench the majority of dimers are in one of the equilibrium buckling states. The final quench values are highlighted as orange symbols. They can be found again in \autoref{Fig::Quench-Result}.}
    	\label{Fig::Process-with-Snapshots}
    \end{figure*}	
    \begin{figure*}
    	\begin{subfigure}{0.48\textwidth}
    		\centering
    		\includegraphics[width=0.95\linewidth]{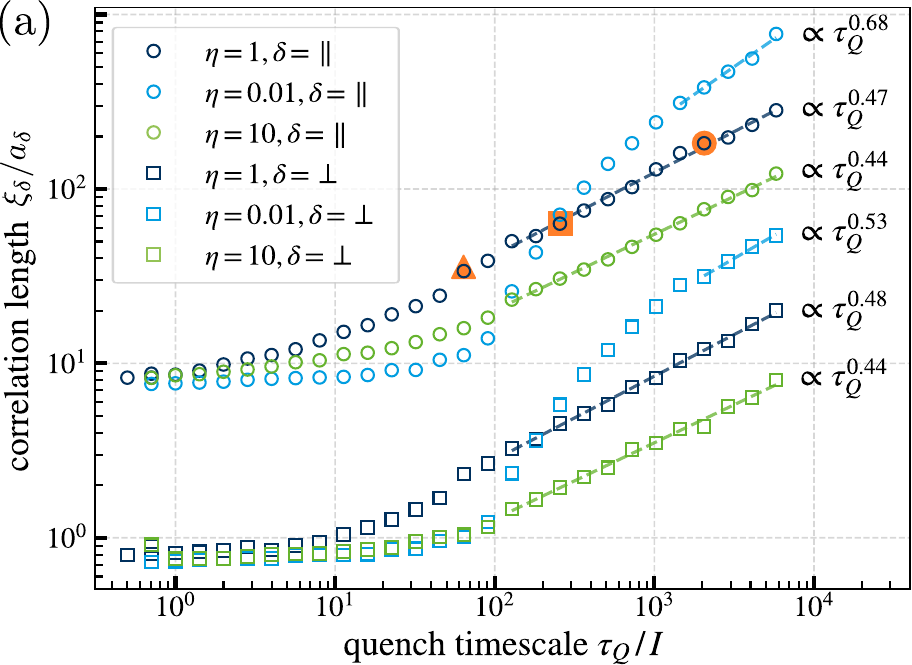}
    	\end{subfigure}
    	\begin{subfigure}{0.48\textwidth}
    		\centering
    		\includegraphics[width=0.95\linewidth]{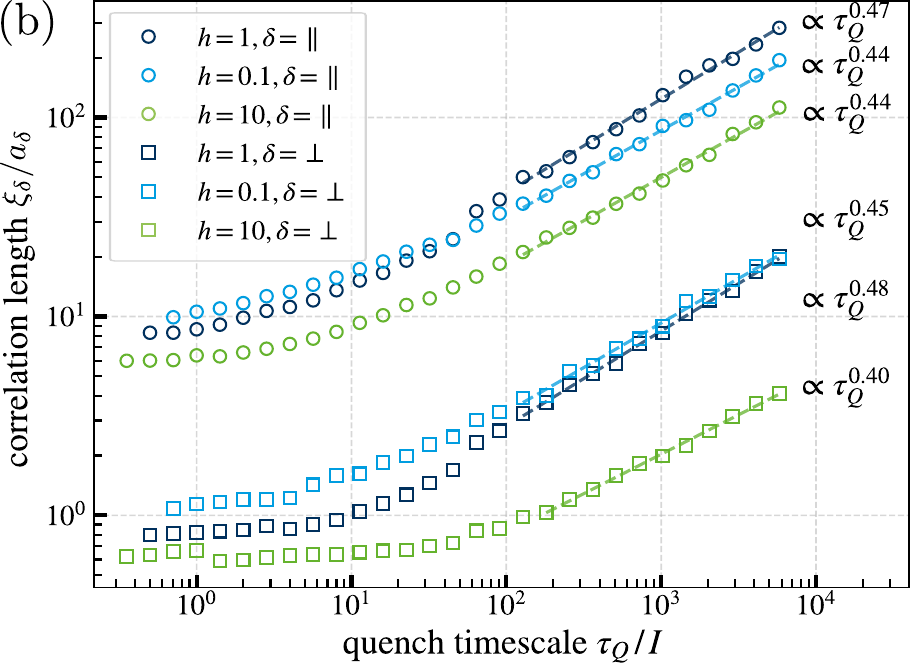}
    	\end{subfigure} 
    	\caption{The quench scaling, i. e., the frozen correlation length after a cooling quench versus the quench timescale is shown. The data poincts corresponding to the final values of \autoref{Fig::Process-with-Snapshots} are highlighted in the left panel. For rapid quench rates the system is unable to evolve adiabatically from the outset. The resulting frozen correlation length corresponds approximately to the equilibrium correlation length at the initial temperature, and no scaling behavior is observed. Hence, the scaling is calculated in the linear area. The extracted quench exponents of $\mu_Q^\parallel = 0.47$ of and $\mu_Q^\perp = 0.48$ are larger than the expected $\mu_{\text{KZM}} \approx 0.32$ for the Ising model. The quench scaling is simulated for systems with varying $\eta$ and $h$ \aw{in panel \textbf{(a)} and \textbf{(b)} respectively}. \aw{Unlike the damping $\eta$ which solely influences the dynamics, the field strength $h$ shifts the critical point (see \autoref{Fig::Phase-Diagram}) and therefore alters the quenched temperature range.} Besides for the cases with small damping, where a bump in the scaling region becomes visible, the quench exponent does not show a clear systematic dependence.}
    	\label{Fig::Quench-Result}
    \end{figure*}
\subsection{Quenches} \label{Section::Quenches}
    We introduce the general  quench exponent $\mu_Q$ by the scaling of the quenched correlation length $\xi$ with the quench timescale $\tau_Q$
	\begin{equation}
		\xi \propto \tau_Q^{\mu_Q}.
	\end{equation}
	The KZM	has been verified in numerous experimental~\cite{ruutu1996vortex, ulm2013observation, pyka2013topological} and numerical~\cite{laguna1997density, schaller2023sequential, antunes2006domain} investigations. Hence, we expect a value of $\mu_Q =	\mu_{\text{KZM}} \approx 0.32$ from previous simulations. We will examine linear quenches like described by Eq.~\eqref{Eq::Linear-Quench} for a temperature range symmetric around the critical point
	\begin{equation}
		\varepsilon_i =	-\varepsilon_f,
	\end{equation}
	so that the absolute values of the initial temperature $\varepsilon_i$ and the final temperature $\varepsilon_f$ match. The simulation is stopped right after $\varepsilon_f$ is reached, so there is no after-quench equilibration phase. \autoref{Fig::Process-with-Snapshots} shows the time resolved quench process in terms of the parallel correlation length $\xi_\parallel(t)$. The absolute value $|\varepsilon_i| = |\varepsilon_f|$ is chosen so that the evolution at the beginning of the quench is approximately adiabatic, meaning the orange equilibrium curve and the dynamic path in \autoref{Fig::Process-with-Snapshots} overlap. At discrete time points, the state of the quenched ensemble is represented by a section snapshot of a quenched system at the corresponding time. The results of the quenches at different timescales are shown in dark blue in \autoref{Fig::Quench-Result}. The calculation of the correlation length is described in \renewcommand{\sectionautorefname}{Appendix} \autoref{Section::Corr-Length-Calculation}\renewcommand{\sectionautorefname}{Section}. The system sizes are chosen so that the frozen correlation length is smaller than $L_\delta/10$.  The extracted quench exponent for $\eta = h = 1$ of $\mu_Q \approx 0.47$ deviates strongly from the expected value $\mu_{\text{KZM}}$. Since the parameters $h$ and $\eta$ have been observed to influence the dynamics drastically, it might be worthwhile to observe their influence on the quench exponent. This is investigated by the other curves in \autoref{Fig::Quench-Result}. Large $\eta$ as well as large $h$ slow down the dynamics of the quench and result in the linear scaling region of $\xi$ moving towards slower quenches. The slope of the scaling region however shows, except for small $\eta$, no sensitive dependence on both parameters. Small $\eta$ reveal the existence of a bump in the $\xi(\tau_Q)$ curve. This might be due to a retardation effect between the controlled reservoir temperature and the state of the system, which is influenced also by the kinetic energy of the dimers. The rotation of the dimers decays slower for small $\eta$ and therefore the system keeps a larger energy per degree of freedom for a longer time. This might lead to the effect that quickly quenched underdamped systems are dominated by the rotational energy. For fast quenches, no significant change in the average dimer energy has happened until the end of the quench. As the quench timescale becomes larger, rotation is damped and the usual mechanisms take over rapidly. With the current results, a dependence of $\mu_Q$ on the quantitative values of $h$ or $\eta$ is improbable. Hence, averaging the quench exponents while neglecting the $\eta = 0.01$ case is sensible. We extract
    \begin{equation}
        		\aw{\mu_{Q} =	0.45\,.}
    \end{equation}
	\aw{The statistical uncertainty calculated as the standard error of \autoref{Fig::Quench-Result} is less than 0.01 and also expected to be smaller than the systematic error.}	An explanation for the deviation to Eq.~\eqref{Eq::KZM-Exponent} might be found in the so called quench angle, defined in \renewcommand{\sectionautorefname}{Appendix}\autoref{Section::Phase-Diagram}\renewcommand{\sectionautorefname}{Section}, which has been introduced by~\cite{mathey2020activating} and~\cite{ladewig2020kibble}. As our system is quenched, it traces a certain path in the three-dimensional phase space spanned by the effective coupling constants. Depending on the quantitative values of the coupling parameters and the quench protocol, the path crosses the phase boundary at a certain angle. This quench angle and the velocity at which the critical point is crossed open up subleading scaling regimes with $\mu_Q>\nu /	(1 + \nu z)$ that are dictated by the irrelevant exponents: The KZM scaling is obtained for very slow quenches or orthogonal quench angles. Further research is required to confirm or deny this hypothesis.
	
	Previously, some of us have simulated~\cite{schaller2023sequential} the quench with the Ising model, for which a quench exponent of $\mu_Q \approx 0.35$ much closer to Kibble-Zurek scaling was found. Additionally, the linear scaling region of $\xi_\perp$ was shifted to larger quench time scales in comparison with $\xi_\parallel$, resulting in a modification of the quenched correlation length ratio $\widehat{\xi}_\parallel\, /	\widehat{\xi}_\perp$. This behavior is also found in the present simulation. The strongest effect is visible in the large-$h$ curves in \autoref{Fig::Quench-Result}. Consequently, the ratio $\widehat{\xi}_\parallel\, /	\widehat{\xi}_\perp$ in the scaling region increases with increasing $h$.
\section{Conclusion} \label{Section::Conclusion}
	During this work, Langevin dynamics methods have been employed to investigate the phase transition on the Si(001) surface. The main innovation has been the step from discrete descriptions using the Ising model to a continuous formulation utilizing an adapted XY model. The purpose of the continuous description is a more realistic modeling of dynamic non-equilibrium behavior. The used Langevin equations were  numerically implemented using parallel processing techniques on GPUs.
	
	Finite-size analysis using the Binder cumulant has determined the static critical exponent $\nu$ and the dynamic exponent of the twofold symmetry-broken XY model to $\nu = 1.04 $ and $z = 2.13$. Strong arguments for the validity of José al.'s work~\cite{jose1977renormalization} for rational $2 < p \leq 4$ have been found. Consequently, the symmetry-broken XY model with rational $p$ is strongly expected to belong to the Ising universality class. The dynamics on the silicon surface have been studied for $J_\parallel /	J_\perp = 100$. The quench exponent has been extracted to $\mu_{Q} =	0.442$, which is significantly larger than the expected Ising value. The influence of the amplitude of the symmetry-breaking field as well as the damping $\eta$ on the quench exponent have been investigated. A dependence of $\mu_Q$ could not be established. The quench angle~\cite{mathey2020activating} is hypothesized as source for the deviation between KZM scaling and the observed scaling.
 
	However, the reason for the deviation is uncertain at the moment. The measurement of as-slow-as-possible quenches might be a starting point for further investigations, although from the current point of view, more than ten times slower quenches will not be realizable. If the deviation from the KZM exponent is caused by the subleading scalings described in~\cite{mathey2020activating, ladewig2020kibble}, another approach could be to quench orthogonally to the phase boundary. The expected KZM scaling should then be observed also for shorter quench times. However, it is not trivial to construct an orthogonal quench protocol because the relevant couplings, that might be combinations of naive $\beta J_\delta, \beta h$, have to be determined first. An insightful extension to the simulation would be to allow nonlinear and anisotropic quenches. According to~\cite{mathey2020activating}, nonlinear quench protocols of higher order $\gamma \in \mathbb{N}$ following $\varepsilon(t) =	t^\gamma /	\tau_Q^\gamma$ might activate scaling regions governed by the smallest irrelevant exponents. This could give valuable hints about the deviations from KZM scaling that we observe.  Anisotropic quenches might be relevant for experimental systems, in which the temperature of the bath and the surface is not perfectly isotropic. 
    Another interesting avenue of further research is the investigation of the parameter $p$. As long as $p \in (2, 4]$, so that the number of minima of the symmetry-breaking field stays the same, it is not expected that the critical behavior changes. However, since it changes the height of the barrier, $p$ is going to have a quantitative influence. Nevertheless, the study of the influence of $p$ on the dynamics might be worthwhile, since the parameter also alters the shape of the symmetry-breaking field.
	
	To conclude, the investigation of the anisotropic XY model with rational $p$ by Langevin dynamics simulations gives a first confirmation of the Ising-like dynamic critical exponent $z$. The investigation of cooling quenches showed nonuniversal behavior, possibly related to the damping or the quench angle.
\section*{Acknowledgement}
    We thank P. Kratzer for fruitful discussion. We also thank the DPG for fundings through the Collaborative Research Center 278162697-SFB 1242. 
\appendix
\section{Numerics} \label{Section::Appendix-Numerics}
\subsection{Integration Method} \label{Section::Integration-Method}
    Many integration schemes have been developed to solve Eqs.~\eqref{Eq::Si-Langevin-theta}, each with their own advantages~\cite{larini2007langevin}. The Brünger-Brooks-Karplus (BBK) method integrates in three steps~\cite{izaguirre2001langevin}: a half-kick (we omit indices for brevity)
	\begin{flalign}
		\quad \omega\big(t + \tfrac{1}{2} \text{d}t \big) =	\big(1&- \tfrac{1}{2} \eta \text{d}t \big ) \omega(t) - \tfrac{1}{2} \text{d}t \tfrac{\partial V(\vartheta(t))}{\partial \vartheta} && \label{Eq::BBK-1}\\
		&+ \tfrac{1}{2} \sqrt{2\eta{k_B T}} n(t) \sqrt{\text{d}t}~, && \nonumber				
	\end{flalign}
	is followed by a drift
	\begin{flalign}
		\quad \vartheta(t + \text{d}t) = \vartheta(t) +  \omega\left(t + \tfrac{1}{2} \text{d}t\right) \text{d}t~,&&
	\end{flalign}
	which is followed by another half-kick
	\begin{flalign}
		\quad \omega\left(t + \text{d}t \right) =	\left(1 + \tfrac{\text{d}t}{2} \eta \right)^{-1} \Big( &\omega(t+ \tfrac{\text{d}t}{2} ) - \tfrac{1}{2} \text{d}t \tfrac{\partial V(\vartheta(t + \text{d}t))}{\partial \vartheta} && \label{Eq::BBK-2} \\
		&+ \tfrac{1}{2} \sqrt{2\eta{k_B T}} n({t + \text{d}t}) \sqrt{\text{d}t} \Big). && \nonumber
	\end{flalign}
	Here, the variable $n(t)$ is a sample value of a Gaussian normal distribution with mean zero and unit standard deviation. The moment of inertia $I$ has been set to one. The GPU implementation makes use of the fact that the calculations Eqs. \eqref{Eq::BBK-1} - \eqref{Eq::BBK-2} on each lattice site can be simultaneously performed in real time, allowing much larger systems to be simulated. Inspired by Ahnert et al.'s architecture of solving differential equations on GPUs~\cite{ahnert2014solving}, our implementation has been extended to accommodate stochastic differential equations. The random numbers $n(t)$ have been generated using CUDA's built-in library cuRAND. \\
	\begin{figure}
		\begin{subfigure}{0.49\textwidth}
			\centering
			\includegraphics[width=0.975\linewidth]{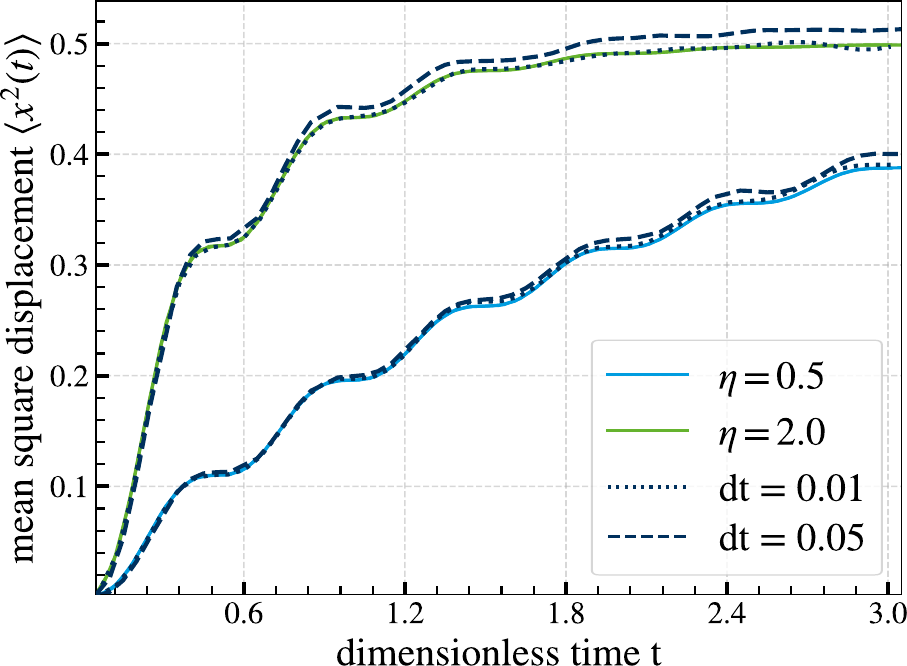}
		\end{subfigure}
		\caption{The calculated $\left \langle x^2 (t) \right \rangle $ of thermal harmonic oscillators are shown for two step sizes. The solid curves are the theoretical paths calculated from Eq.~\eqref{Eq::MSD-HO} for $\omega^2 =	20$, $T	=	20$ and $m=1$.	The damping is denoted in the legend. The dotted lines are paths calculated from $10^5$ simulated trajectories. The BBK method in this use case is very stable and shows only small discretization errors up to a step size of $\text{d}t = 0.05$. For a step size of $\text{d}t = 0.01$, virtually no discretization errors occur.}
		\label{Fig::MSD-Comparison}
	\end{figure}
\subsection{Benchmarks} \label{Section::Benchmarks}
    Since an analytic solution of our model is intractable, proper benchmarks are vital to ensure the correctness of our simulation. Furthermore, the step size $\text{d}t$ used to integrate the Langevin equations has to be analyzed to ensure convergence as well as avoiding discretization errors while being efficient. The benchmarks that were conducted are the dynamics of independent harmonic oscillators coupled to a thermal reservoir, and the equilibrium distribution of particle pairs described by an Hamiltonian analogous to Eq.~\eqref{Eq::XY-Hamilton-Kinetic}. The equation that describes the time evolution of probability densities of Brownian motion is the \textit{Fokker-Planck equation}. When talking about the probability density $p(x, v, t)$ in terms of particle velocity $v$ and position $x$, it is often referred to as the \textit{Klein-Kramers-} or \textit{Smoluchowski} equation and written as
	\begin{equation} \label{Eq::Klein-Kramers}
		\begin{split}
			\frac{\partial}{\partial t} p(x, v, t) = \Bigg(-\frac{\partial}{\partial x} v &+ \frac{\partial}{\partial v} \Big [\eta v - \frac{1}{m} \frac{\partial}{\partial x} V(x) \Big] \\
			&+ \frac{\eta k_B T}{m} \frac{\partial}{\partial v^2}\Bigg)p(x, v, t)~.
		\end{split}
	\end{equation}
    The Fokker-Planck equation and the Langevin Eqs.~\eqref{Eq::Si-Langevin-theta} are equivalent~\cite{risken1996fokker}. The distribution of paths following Langevin equations results in the probability distribution satisfying Eq.~\eqref{Eq::Klein-Kramers}. The steady state distribution of the Fokker-Planck equation is the canonical distribution
	\begin{equation} \label{Eq::Canonical-Dist}
		p(x, v) \propto e^{-\beta \left( \tfrac{1}{2} m v^2 + V(x)\right)},
	\end{equation}
	allowing for a simple way to verify long term behavior. 

    For uncoupled harmonic oscillators subject to a 
    quadratic potential
	\begin{equation}
		V(x) =	\tfrac{1}{2} \omega^2 x^2,
	\end{equation}
	the Fokker-Planck equation is analytically solvable~\cite{risken1996fokker}. This makes it suitable to check correct dynamics of the simulation. The analytic solution for the second moment of $x(t)$ reads
	\begin{equation} \label{Eq::MSD-HO}
		\begin{split}
			\left \langle x^2 \right \rangle (t)  =&	\frac{\eta {k_B T}}{m (\lambda_+ - \lambda_-)^2} \bigg[ \frac{\lambda_+ + \lambda_-}{\lambda_+ \lambda_-} \\
			&+ \frac{4}{\lambda_+ + \lambda_-} \left(e^{- (\lambda_+ + \lambda_-) t} - 1\right) \\
			&- \frac{1}{\lambda_+} e^{-2\lambda_+ t} - \frac{1}{\lambda_-} e^{- 2 \lambda_- t}\bigg],
		\end{split}
	\end{equation}
	with
	\begin{equation}
		\lambda_{\pm} =	\frac{1}{2} \left(\eta \pm \sqrt{\eta^2 - 4 \omega^2}\right)~.
	\end{equation}
	In \autoref{Fig::MSD-Comparison} we compare $\left \langle x^2 \right \rangle (t)$ calculated from $10^5$ paths, simulated by the BBK method, with the theoretical result. The algorithm with $\text{d}t=0.05$ yields small deviations from the theoretical curve. For a smaller step size of $\text{d}t = 0.01$, the accuracy of the method for thermal oscillators is much better. 
    
    The second benchmark, consisting of two interacting particles in a cosine potential, has the purpose of verifying the correct behavior of the interaction as well as thermalization. We consider the one-dimensional version of Eq.~\eqref{Eq::XY-Hamilton-Kinetic} for two sites, so particle pairs in a cosine potential interacting via an XY interaction. \aw{Analogous to the actual simulation, the state space is restricted to $\vartheta \in [-\pi /	2, \pi /	2)$ and the potential is chosen as in \hyperref[Fig::States]{Figure 3b}}. The probability distribution becomes a four-dimensional function $p(\vartheta_1, \vartheta_2, \omega_1, \omega_2)$. We expect that this distribution relaxes to the equilibrium distribution of the Fokker-Planck equation, \aw{i.e.} the canonical distribution. 
    
\begin{figure*}[htp]
		\begin{subfigure}{0.325\textwidth}
			\centering
			\includegraphics[width=0.98\linewidth]{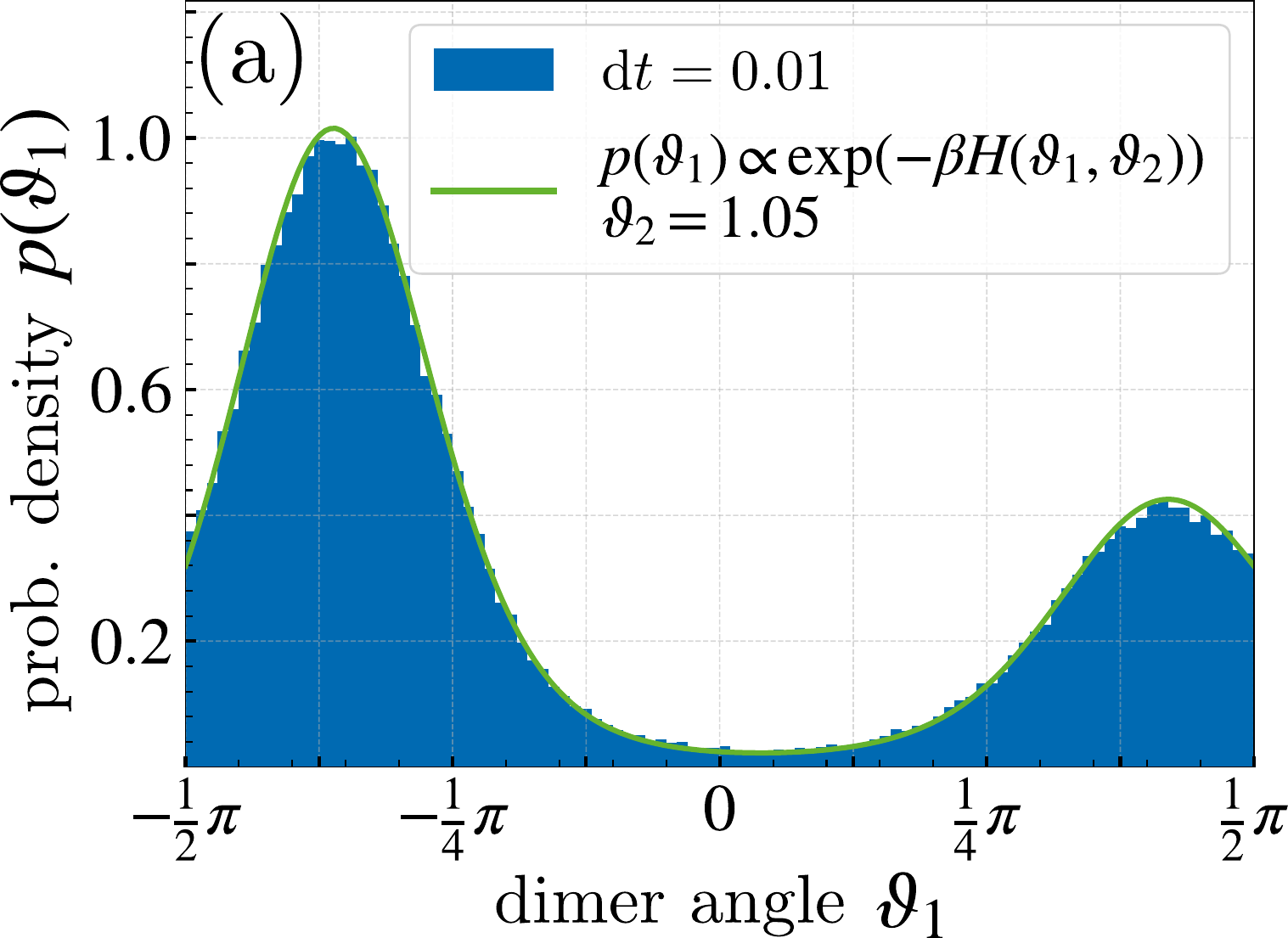}
		\end{subfigure}
		\begin{subfigure}{0.325\textwidth}
			\centering
			\includegraphics[width=0.98\linewidth]{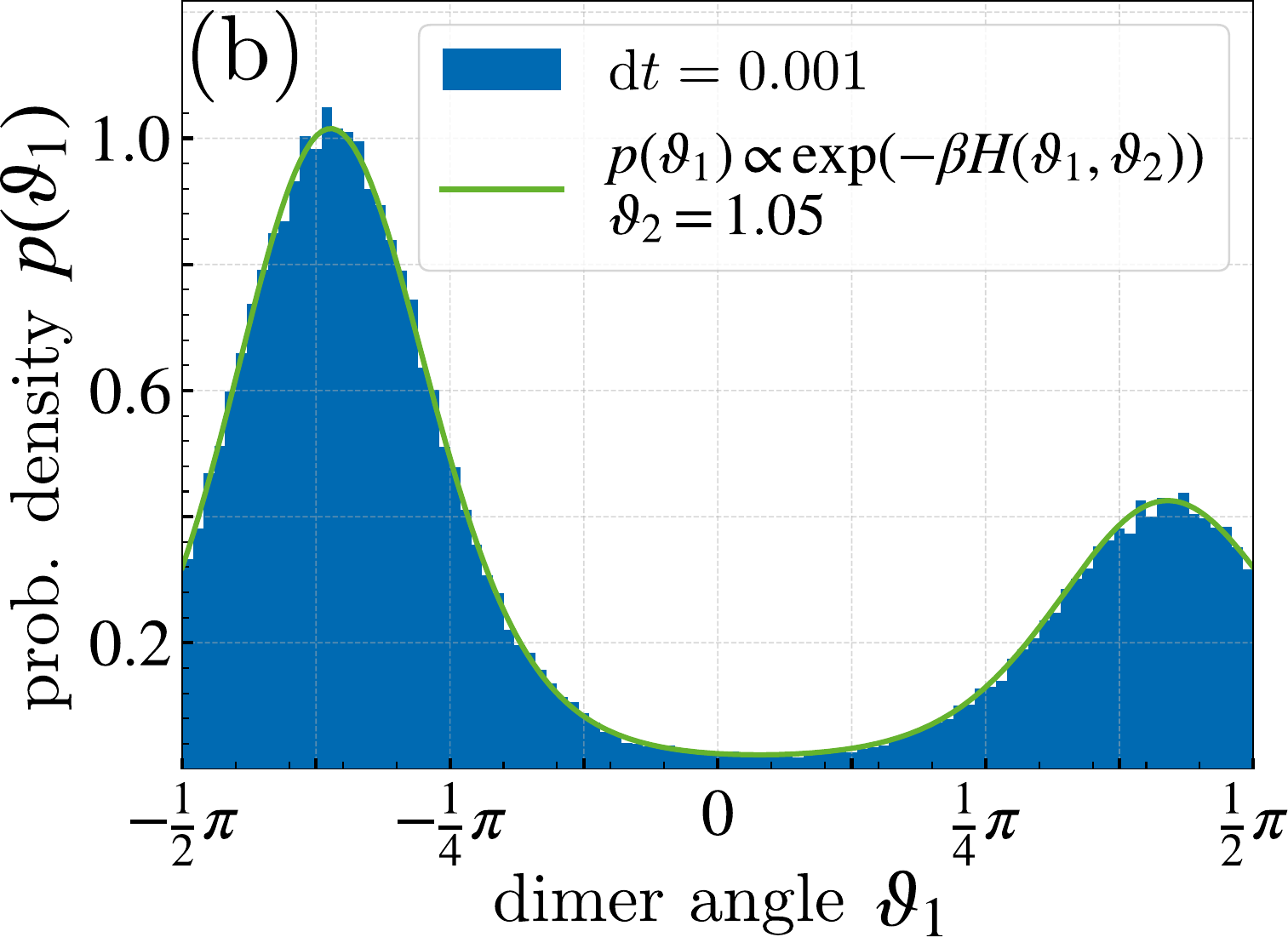}
		\end{subfigure}
		\begin{subfigure}{0.325\textwidth}
			\centering
			\includegraphics[width=0.98\linewidth]{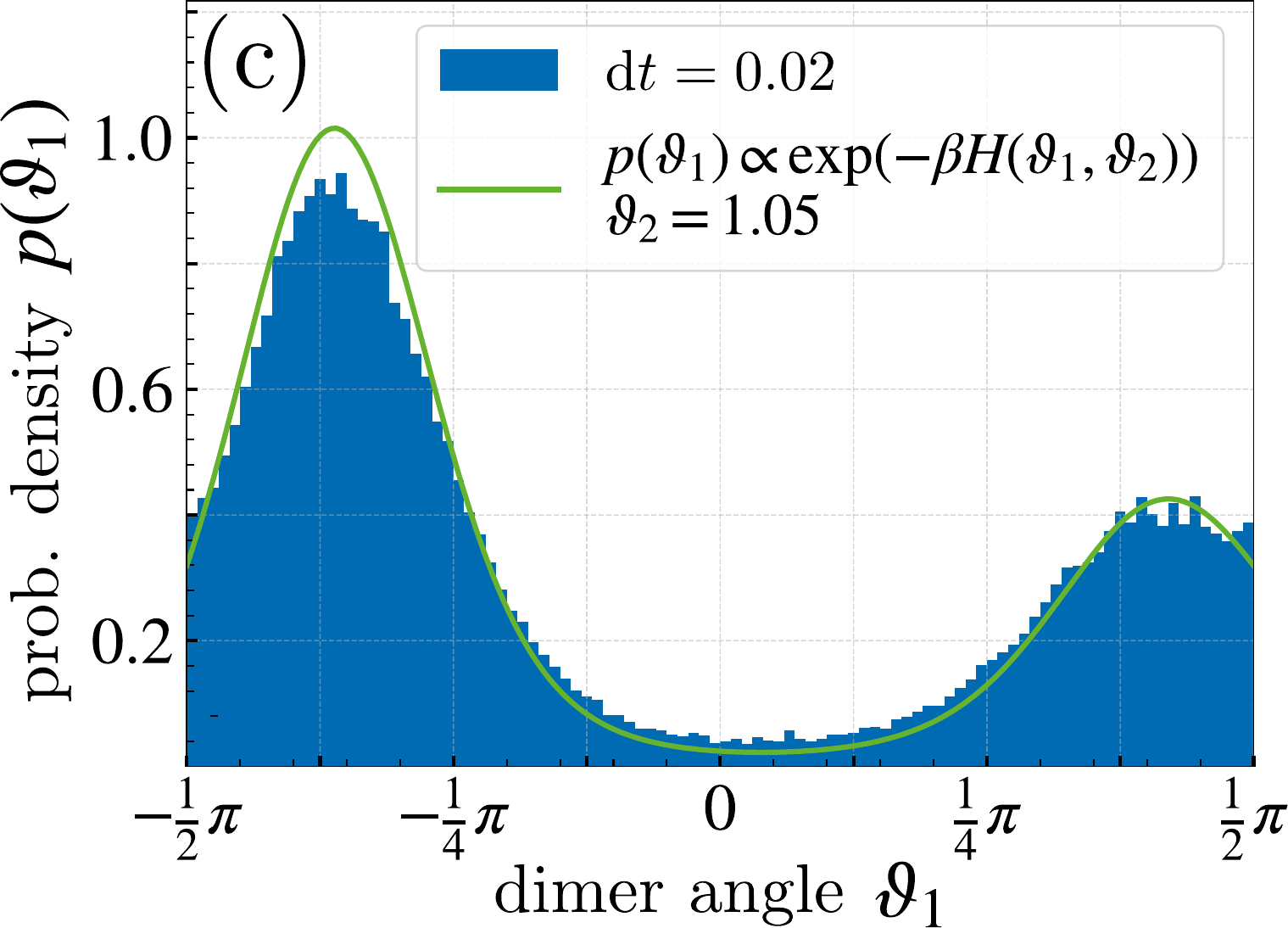}
		\end{subfigure}
		\caption{The integrated probability distribution $p(\vartheta_1) = p(\vartheta_1, \vartheta_2=1.05)\Delta\vartheta_2$ of particle pairs described by Eq.~\eqref{Eq::XY-Hamilton-Field} around constant $\vartheta_2$ with $\Delta \vartheta_2 = 0.15$ is shown. The used parameters for (a) - (c) were $J =	40/I,~ h =	100 /	I,~ p=2.5,~ \eta =	1$ and $T= 100 /	I$. The interactions and temperatures are chosen to be larger than in the actual simulation on purpose. The green curve is the canonical distribution. The blue bins are calculated from $5 \cdot 10^5$ simulated pair paths by counting particles in $\left[\vartheta_2 - \tfrac{1}{2} \Delta\vartheta_2, \vartheta_2 + \tfrac{1}{2} \Delta\vartheta_2 \right]$ and binning them into $100$ sections for $\vartheta_1$. It was made sure that the systems were completely relaxed, meaning that the effects of the starting position vanished and the shape of the probability distribution was stable. The BBK method again shows no discretization errors for $\text{d}t=0.01$. Slightly larger step sizes of $\text{d}t =	0.02$ already produce visually recognizable deviations.  Almost no improvement is achieved by reducing the step size to $\text{d}t =	0.001$.}
		\label{Fig::Pair-Prob-Dist}
	\end{figure*}

    A suitable representation is achieved by integrating out the angular velocities and considering a cross section of a fixed interval $\left[\vartheta_2 - \tfrac{1}{2} \Delta\vartheta_2, \vartheta_2 + \tfrac{1}{2} \Delta\vartheta_2 \right]$.
	In \autoref{Fig::Pair-Prob-Dist} we show cross sections of the integrated probability density 
	\begin{equation}
		p(\vartheta_1, \vartheta_2) \Delta\vartheta_2 =	\int_{}\text{d}\omega_1 \,\text{d} \omega_2\, p(\vartheta_1, \vartheta_2, \omega_1, \omega_2) \Delta\vartheta_2,
	\end{equation}
	with constant $\vartheta_2$. It is verified that simple interacting systems are driven to their thermal equilibrium. The statistical and discretization errors vanish for many samples and small step sizes $\text{d}t$.	

\section{Extraction of observables}
	\label{Section::observable-measurement}
	The observables of interest are usually obtained by \textit{ensemble averages} of systems in {thermal equilibrium}. The ensemble average is calculated by computing the desired observable for many different realizations of the same system.  The \textit{ergodic hypothesis} assumes that the average of many realizations and the average over one long-term simulation are the same. We assume that our system is ergodic. Since GPU accelerated programming is employed, we want to harness the full potential of the graphical processing unit. Systems of sufficient scale should be simulated to optimize GPU usage. Depending on the GPU, maximum performance is usually acquired by using systems with more than $N=  5 \cdot 10^5...\,1 \cdot 10^6$ lattice sites. For many use cases, e.g. calculating the Binder cumulant $U$ in \autoref{Section::Static-scaling}, systems of this size are not required. Therefore, always a number $n \approx	N /	(L_\parallel L_\perp)$ of independent subsystems is simulated. For a simulation of $n$ subsystems over a time $\tau_s$, the ensemble average of a quantity $f$ is calculated via
	\begin{equation} \label{Eq::ensemble-average-calculation}
		\langle f \rangle = \frac{1}{n \cdot \tau_s} \sum_i^n \int_{t_0}^{t_0 + \tau_s} \text{d}s f_i(s)~.
	\end{equation}
    To calculate the Binder cumulant Eq.~\eqref{Eq::Def-Binder-Cum}, the ensemble averages of $f = M^2$ and $f = M^4$ have to be evaluated. In case of the correlation length extraction relevant in \autoref{Section::Quenches} and described in \renewcommand{\sectionautorefname}{Appendix}\autoref{Section::Corr-Length-Calculation}\renewcommand{\sectionautorefname}{Section}, the Fourier transform~\eqref{Eq::FT-Corr-delta} of the correlation function $S_\delta(k)$ is calculated as the summed squares of the lattice Fourier transform~\eqref{Eq::S-as-lattice-FT} and averaged before fitting Eq.~\eqref{Eq::Lorentzian-Peak}. For the time-dependent Binder cumulant $U(t, \varepsilon, L)$ of \autoref{Section::dynamic-scaling}, the time integration in Eq.~\eqref{Eq::ensemble-average-calculation} is restricted to an interval $s \in \big[t - \tfrac{\text{d}\sigma}{2} , t + \tfrac{\text{d}\sigma}{2}\big)$ with the step size $\text{d}\sigma$ chosen for suitable representation. The averaging process is repeated until major fluctuations have averaged out. When making use of the ergodic hypothesis to calculate equilibrium quantities, like in \autoref{Fig::Binder-Cum-Result} or the orange curve in \autoref{Fig::Process-with-Snapshots}, it is important to make sure that the system is actually thermalized. Usually, a lower time bound $t_0$ is introduced to account for the relaxation time. Again, the relaxation time is not known and therefore, $t_0$ is set to a fraction of $\tau_s$ to eliminate strong fluctuations in the equilibration phase of the simulation. To eventually judge the relaxation of the system, the uncertainty of $\langle f \rangle$ is considered. The approach is that the equilibration introduces a large statistical deviation on the average $\langle f \rangle$. So by calculating the error $\Delta \langle f \rangle $, we can extract information about the state of equilibration of the system. The $f_i(s)$ are correlated for different points in time. The strength of the correlation depends on how quickly the system relaxes. To get reasonable estimates on the error of $\langle f \rangle$, we have to judge how many effectively independent readings of $f_i$ are taken by the $\text{d}s$ integration in Eq.~\eqref{Eq::ensemble-average-calculation}. Analyzing $f_i(s)$ as time series helps to avoid terminating the simulation too early for slowly relaxing systems. Additionally, it gives an estimate for how many measurements should be recorded in order not to overflow with data. \\
	
	On the run, $f_i(s)$ is calculated and saved for discrete time steps $\text{d}s$. The sample time step $\text{d}s$ is usually much larger than the integration step size
	\begin{equation}
		\text{d}s \gtrsim 100 \cdot \text{d}t,
	\end{equation}
	to limit the amount of data generation. The exact value depends on the integrated autocorrelation time $\tau_C$ \eqref{Eq::Autocorrelation-Time} and is adapted during the simulation
	\begin{equation}
		\text{d}s \approx \frac{1}{10} \tau_C~.
	\end{equation}	
	An average that is calculated by the means of Eq.~\eqref{Eq::ensemble-average-calculation} has a non-trivial relationship with its variance~\cite{madras1988pivot}. The reason for this is that $f_i(s)$ and $f_i(s + \text{d}s)$ can be correlated. Observables at different points in time are dependent on each other  and have to be treated accordingly. The average $f_{\tau_s}$ of a time series $f(s)$ (omit indices for brevity) is calculated as
	\begin{equation}
		f_{\tau_s}  = 	\frac{1}{\tau_s} \int_0^{\tau_s} \text{d}s f(s)~,
	\end{equation}
	with $\tau_s$ being the duration of the sampling. To estimate the error on $f_{\tau_s}$ we consider its variance~\cite{frenkel2023understanding, anderson2011statistical}
	\begin{equation} \label{Eq::Time-Series-Var}
		\begin{split}
			\sigma_{f_{\tau_s}}^2 &=	\left \langle f_{\tau_s}^2 \right \rangle - \left \langle f_{\tau_s} \right \rangle^2 \\
			&\approx \frac{1}{\tau_s} \int_{-\infty}^{\infty} \text{d}t~C_f(t),
		\end{split}
	\end{equation}
	with $C_f(t)$ being the autocorrelation or time correlation function
	\begin{equation} \label{Eq::Autocorrelation-Function}
		C_f(t) =	\lim_{s \rightarrow \infty} \left( \left \langle f(s) f(s + t) \right \rangle - \left \langle f(s) \right \rangle^2 \right)~.
	\end{equation}
	The step performed in Eq.~\eqref{Eq::Time-Series-Var} is valid in the limit that the sampling time is much larger than the characteristic decay time of the autocorrelation function $\tau_C$. The so called \textit{integrated autocorrelation time} $\tau_C$ is defined as
	\begin{equation} \label{Eq::Autocorrelation-Time}
		\tau_C = \frac{1}{2}	\int_{-\infty}^{\infty} \text{d}t~C_f(t) /	C_f(0)~.
	\end{equation}
	The variance $\sigma_{f_{\tau_s}}^2$ is expressed in terms of $\tau_C$ as
	\begin{equation}
		\sigma_{f_{\tau_s}}^2 =	\frac{2\tau_C}{\tau_s} C_f(0)~.
	\end{equation}
	Consideration of Eq.~\eqref{Eq::Autocorrelation-Function} shows that $C_f(0)$ reduces to the variance of $f$
	\begin{equation}
		C_f(0) =	\sigma_f^2~.
	\end{equation}
	Rewriting $\tau_s =	n_s \Delta s$ with $n_s$ being the number of measured samples and $\text{d}s \rightarrow \Delta s$ being the discrete time between the samples, the variance of the mean $f_{\tau_s}$ can be expressed as
	\begin{equation}
		\sigma_{f_{\tau_s}}^2 =	\frac{2 \tau_C}{\tau_s}{\sigma_f^2} = \frac{2 \tau_C}{\Delta s} \frac{\sigma_f^2}{n_s}~,
	\end{equation}
	revealing that the variance of $f_{\tau_s}$ is by a factor of ${2 \tau_c} /	{\Delta s}$ larger than the naive approach of uncorrelated measurements. Since it is practically not possible to integrate Eq.~\eqref{Eq::Autocorrelation-Time} from $-\infty$ to $\infty$,   we approximate $\tau_c$ by
	\begin{equation}
		\tau_C \approx \frac{1}{2} \int_{-\tau_s/2}^{\tau_s/2} \text{d}t~C_f(t) /	C_f(0)~.
	\end{equation}
\section{Correlation Length Extraction} \label{Section::Corr-Length-Calculation}
	The two-point equal time correlation function of the XY model with rotors $\vec{s}_{x, y} = (\cos \vartheta_{x, y}, \sin \vartheta_{x, y})$ in two dimensions is defined as
	\begin{equation}
		C(x, y) = \langle \vec{s}_{0,0} \vec{s}_{x, y} \rangle~.
	\end{equation}
	The brackets $\langle~\cdot~\rangle$ denote the ensemble average
	\begin{equation}
		\langle \vec{s}_{0,0} \vec{s}_{x, y} \rangle  = \frac{1}{Z} \int \prod_i d\vartheta_i \vec{s}_{0,0} \vec{s}_{x, y} e^{- \beta H(\{\vartheta\})}
	\end{equation}	
	The correlation function decays exponentially for large distances above the critical temperature~\cite{kosterlitz1974critical, amit1980renormalisation}, following	 
	\begin{equation} \label{Eq::corr-func-decay-above-Tc}
		C(x, y) \sim e^{-r(x,y) /	\xi(x,y)} \quad \text{with} \quad r(x,y) =	\sqrt{x^2 + y^2}.
	\end{equation}
	This is the definition of the correlation length $\xi$. The correlation length is a measure for the length scale over which perturbations of a system relax in space. We are mainly interested in the correlation lengths in the directions along and across the dimer row and therefore define the correlation functions in these directions as
	\begin{align} \label{Eq::Corr-Func-asymptotic}
		\begin{split}
			&C_\perp(y) =  \langle \vec{s}_{0,0} \vec{s}_{0, y} \rangle \sim e^{-y /	\xi_\perp} \qquad \text{and} \\
			&C_\parallel(x) =  \langle \vec{s}_{0,0} \vec{s}_{x, 0} \rangle \sim e^{-x /	\xi_\parallel}.
		\end{split}
	\end{align}
	Consider the Fourier transforms of $C_\delta(r)$, $\delta \in \lbrace \parallel \,,\, \perp \rbrace$,
	\begin{equation}  \label{Eq::FT-Corr-delta}
		S_\delta(k) = \sum_{r=0}^{N_\delta - 1} C_\delta (r) e^{-2\pi \mathrm{i} \frac{kr}{N_\delta}}~,
	\end{equation}
	with $N_\delta$ being the number of lattice sites in the direction of $\delta$. Set in the following $\delta =\,	\perp$. The Fourier transform becomes 
	\begin{equation} \label{Eq::FT-of-Corr-perp}
		\begin{split}
			S_\perp(k) =~ &\sum_{y = 0}^{N_\perp - 1} C_\perp (y) e^{-2\pi \mathrm{i} \frac{ky}{N_\perp}} =\sum_{y = 0}^{N_\perp - 1} \langle \vec{s}_{0,0} \vec{s}_{0, y} \rangle e^{-2\pi \mathrm{i} \frac{ky}{N_\perp}}\\
			 =~ & \sum_{\kappa = 0}^1 \sum_{y = 0}^{N_\perp - 1} \langle s^\kappa_{0,0} s^\kappa_{0, y} \rangle e^{-2\pi \mathrm{i} \frac{ky}{N_\perp}}, 
		\end{split}
	\end{equation}
    with $\kappa \in \lbrace 0, 1 \rbrace$ denoting the spin component. The ensemble average can be computed by a sum over an infinite lattice
	\begin{equation} \label{Eq::Ensemble-Avg-as-mean}
		\langle s^\kappa_{0,0} s_{0, y}^\kappa \rangle =	\lim\limits_{N_\perp \rightarrow \infty} \lim\limits_{N_\parallel \rightarrow \infty} \frac{1}{N_\perp N_\parallel} \sum_{i =	0}^{N_\perp} \sum_{j=0}^{N_\parallel}   s^\kappa_{i,j} s_{i, j  + y}^\kappa~.
	\end{equation}
    An approximation is possible by using a finite lattice with large dimensions $N_\delta$. Inserting Eq.~\eqref{Eq::Ensemble-Avg-as-mean} into Eq.~\eqref{Eq::FT-of-Corr-perp} and replacing $j + y \rightarrow q$ yields
	\begin{equation} \label{Eq::FT-Corr-Delta}
		\begin{split}
			S_\perp(k) &= \frac{1}{N_\perp N_\parallel}  \sum_{\kappa,q,i,j}^{}   s^\kappa_{i,j} s_{i, q}^\kappa e^{-2\pi \mathrm{i} \frac{k(q-j)}{N_\perp}} \\
			&=	\frac{1}{N_\perp N_\parallel}  \sum_{\kappa,q,i,j}^{}  \left(\sum_{p=0}^{N_\parallel} \delta_{i,p} \right) s^\kappa_{i,j} s_{p, q}^\kappa e^{-2\pi \mathrm{i} \frac{k(q-j)}{N_\perp}}~.
		\end{split}
	\end{equation}
	In the second step we have inserted an identity in the form of  a sum over a Kronecker delta. The Kronecker delta can be written as a sum over complex exponentials
	\begin{equation}
		\delta_{p,j} =	\frac{1}{N_\parallel} \sum_{l=1}^{N_\parallel} e^{2 \pi \mathrm{i} \frac{l(j - p)}{N_\parallel}}~.
	\end{equation}
	Inserting this representation into Eq.~\eqref{Eq::FT-Corr-Delta} gives
	\begin{equation}
		\begin{split}
			S_\perp(k) &=	\frac{1}{N_\perp N_\parallel^2} \sum_l \sum_\kappa \sum_{q,p,i,j} s^\kappa_{i,j} s_{p, q}^\kappa e^{-2\pi \mathrm{i} \frac{k(q-j)}{N_\perp}} e^{2 \pi \mathrm{i} \frac{l(p - i)}{N_\parallel}} \\
			&=	\frac{1}{N_\perp N_\parallel^2} \sum_l \sum_\kappa \Bigg(\sum_{i,j} s^\kappa_{i,j} e^{2\pi \mathrm{i} \left(\frac{kj}{N_\perp} + \frac{li}{N_\parallel} \right)} \Bigg) \\
			 &\qquad\qquad \qquad \times \left(\sum_{q,p} s_{p, q}^\kappa e^{-2 \pi \mathrm{i} \left( \frac{kq}{N_\perp} + \frac{lp}{N_\parallel} \right)} \right)~.
		\end{split}
	\end{equation}
	The expressions in the parentheses are the Fourier transforms $\tilde{s}_{k,l}^\kappa$, or respectively the conjugated Fourier transform, of the lattice $s_{i,j}^\kappa$:
	\begin{equation} \label{Eq::S-as-lattice-FT}
		\begin{split}
			S_\perp(k) &=	\frac{1}{N_\perp N_\parallel^2} \sum_l \sum_\kappa \left(\tilde{s}_{k,l}^\kappa\right)^* \tilde{s}_{k,l}^\kappa \\
			&= \frac{1}{N_\perp N_\parallel^2} \left( \sum_l |\tilde{s}_{k,l}^0|^2  + \sum_l |\tilde{s}_{k,l}^1|^2\right)~.
		\end{split}
	\end{equation}
	This way we can calculate $S_\perp(k)$ by computing the two dimensional Fourier transforms of the lattices $s_{i,j}^0 = \cos \vartheta_{i,j}	$ and $s_{i,j}^1 =	\sin \vartheta_{i,j}$.
	The analogue result is valid for $S_\parallel(k)$. \\
	
	To eventually extract the correlation length, we consider again Eq.~\eqref{Eq::FT-Corr-Delta} and insert the asymptotic behavior of $C_\delta (r)$  Eq.~\eqref{Eq::Corr-Func-asymptotic} to obtain
	\begin{equation} \label{Eq::Lorentzian-Peak}
		S_\delta(k) \sim \sum_r^{N_\delta - 1} e^{-|r| /	\xi_\delta} e^{-2\pi \mathrm{i} \frac{kr}{N_\delta}} = \frac{2 \xi_\delta}{1 + 4 \pi^2 \xi_\delta^2 k^2}	~,
	\end{equation}
	showing that $S_\delta(k)$ behaves like a Lorentzian function around $k=0$. Calculating $S_\delta(k)$ by means of Eq.~\eqref{Eq::S-as-lattice-FT} and fitting to the Lorentzian Eq.~\eqref{Eq::Lorentzian-Peak} yields $\xi_\delta$ as fitting parameter. Since Eq.~\eqref{Eq::corr-func-decay-above-Tc} is valid for large $r$, it is important to only fit the peak of $S_\delta(k)$ around $k = 0$. Correlations below the critical temperature in the Ising universality class can be shown~\cite{mccoy1973two} to decay like
	\begin{equation}
		C(x, y) \sim  c + e^{-r(x,y) /	\xi(x,y)},
	\end{equation}
	with a positive constant $c$. When performing the Fourier transform, this constant contributes to a delta peak at $k = 0$. To correctly extract correlation lengths below $T_c$, the $S_\delta(k =	0)$ value has to be cut.	
\section{Phase Diagram} \label{Section::Phase-Diagram}
\begin{figure}
        \centering        \includegraphics[width=0.95\linewidth]{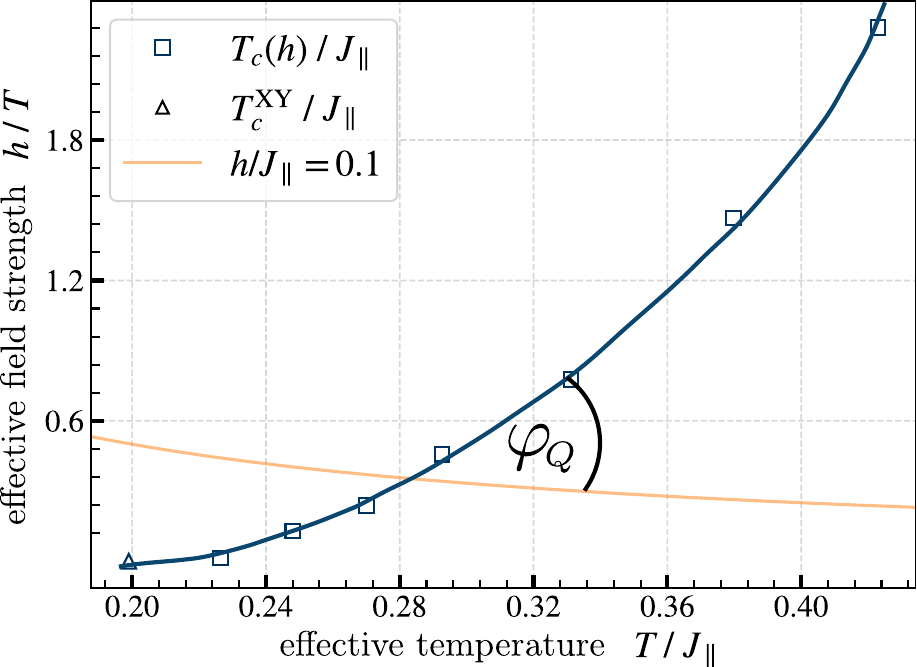}
        \caption{Some simulated critical points are shown for the adapted XY model with $J_\parallel / J_\perp \approx 30$. We added guide for the eye for the phase boundary. The triangle marks the Kosterlitz-Thouless transition of the XY model without symmetry-breaking field calculated by an approximation~\cite{mattis1984transfer}. A system with constant couplings traces a path (orange) in the phase space when cooled down. The path crosses the phase boundary at an angle $\varphi_Q$, in this work called the quench angle. In~\cite{mathey2020activating, ladewig2020kibble} it has been reported that steep quench angles and fast quenches open up subleading scaling regimes that are dictated by irrelevant exponents. }
        \label{Fig::Phase-Diagram}
    \end{figure}

    The critical point of the symmetry-broken XY model depends on the field strength $h$. A cross section of the phase space in the $\beta h$-$(\beta J_\parallel)^{-1}$-plane is shown in \autoref{Fig::Phase-Diagram}. When reducing the temperature while keeping the couplings constant, the system traces a path in the phase space and crosses the phase boundary at the critical temperature. We call the angle between the path and the phase boundary \textit{quench angle}.
    \newline
 \section{Ising Dynamic Critical Exponent} \label{Section::Ising-Crit-Exp}
    The anisotropic Ising model is part of \textit{Model A} as specified by Hohenberg and Halperin~\cite{hohenberg1977theory}. Its dynamic critical exponent $z$ can be expressed in terms of
    \begin{equation}
        z = 2 + c \tilde{\eta}
    \end{equation}
    with the known critical exponent $\tilde{\eta}$ and a constant $c$ to be determined. Some recent results of $z$ are given in \autoref{Tab::Ising-z-values}.
	\begin{table}
		\centering
		\caption{Recent results for the dynamic critical exponent $z$ are summarized. The estimates are obtained by Monte-Carlo methods (MC)~\cite{nightingale2000monte}, renormalization group calculations (RG)~\cite{adzhemyan2022dynamic, duclut2017frequency} and high-temperature expansion (HT)~\cite{dammann1993dynamical}.}
		\renewcommand{\arraystretch}{1.3}
		\begin{tabular}{c c l}
			Source & Method  & Result \\
			\midrule
			\cite{nightingale2000monte} & MC &  $2.167$ \\
			\cite{adzhemyan2022dynamic} & RG &  $2.14 \pm 0.02$ \\
			\cite{duclut2017frequency} & RG &  $2.183 \pm 0.005$ \\
			\cite{dammann1993dynamical} & HT &  $2.15$ \\
			\bottomrule
			$z_\text{Ising}$ & Average &  $2.16$ \\
		\end{tabular}
		\label{Tab::Ising-z-values}
	\end{table}


\newpage


\bibliographystyle{unsrt}
\bibliography{references}
 
\end{document}